\documentstyle[11pt]{article}
\newcommand{\sect}[1] {\section{#1}\setcounter{equation}{0}}
\newcommand{\subsect}[1] {\subsection{#1}
                          \setcounter{equation}{0}}
\renewcommand{\theequation} {\thesection.\arabic{equation}}

\newcommand{\be}{\begin{equation}}
\newcommand{\BE}{\begin{equation}}
\newcommand{\BL}[1] {\begin{equation}\label{#1}}
\newcommand{\EE}{\end{equation}}
\newcommand{\BA}[1] {\begin{eqnarray}\label{#1}}
\newcommand{\EEA}{\end{eqnarray}}
\newcommand{\BAN} {\begin{eqnarray}\nonumber}
\newcommand{\bay}{\begin{array}}
\newcommand{\eay}{\end{array}}

\newcommand{\al}   {\alpha}
\newcommand{\bt}   {\beta}
\newcommand{\ga}   {\gamma}
\newcommand{\ff}   {\varphi}
\newcommand{\dl}   {\delta}
\newcommand{\dz}   {\partial}

\newcommand{\ve}   {\varepsilon}
\newcommand{\lm}   {\lambda}
\newcommand{\sg}   {\sigma}

\newcommand{\la}   {\lambda_a}

\newcommand{\ma}   {\mu_a}
\newcommand{\mb}   {\mu_b}

\newcommand{\ea}  {\varepsilon_a}
\newcommand{\eb}  {\varepsilon_b}
\newcommand{\ala}  {\alpha_a}
\newcommand{\alb}  {\alpha_b}
\newcommand{\bta}  {\beta_a}
\newcommand{\btb}  {\beta_b}
\newcommand{\di}  {\partial_i}
\newcommand{\da}  {\partial_a}

\newcommand{\ppm} {\psi^{\mu}}
\newcommand{\pma} {\psi^{\mu-\varepsilon_a}}
\newcommand{\ppa} {\psi^{\mu+\varepsilon_a}}
\newcommand{\ppi} {P^{l-\varepsilon_i}}

\newcommand{\iaa} {A_{ia}}

\newcommand{\sij} {\sum_{i\ne j}}
\newcommand{\sba} {\sum_{b(\ne a)}}
\newcommand{\suma}{\sum_{|\alpha|=s-r}}
\newcommand{\sumam}{\sum_{|\alpha|=s-r-1}}
\newcommand{\sumi}{\sum_{i,i'}}
\newcommand{\chosa}{{\sigma\choose\alpha}}

\newcommand{\iiss} {I^{(s,s')}}

\newcommand{\ux}   {\underline x}
\newcommand{\uz}   {\underline z}

\newcommand{\uJ}   {\underline J}

\newcommand{\ia}  {(u_i-z_a)}
\newcommand{\ib}  {(u_i-z_b)}
\newcommand{\ja}  {(u_j-z_a)}
\newcommand{\jb}  {(u_j-z_b)}
\newcommand{\ij}  {u_{ij}}
\newcommand{\iiq} {(u_i-v_{i'})}
\newcommand{\iqa} {(v_{i'}-z_a)}
\newcommand{\iqb} {(v_{i'}-z_b)}
\newcommand{\zab} {(z_a-z_b)}

\newcommand{\fia} {\frac 1{(u_i-z_a)}}
\newcommand{\fja} {\frac 1{(u_j-z_a)}}
\newcommand{\fib} {\frac{1}{(u_i - z_b)}}

\newcommand{\fij} {\frac 1{(u_i-u_j)}}
\newcommand{\fiiq} {\frac 1{(u_i-v_{i'})}}

\newcommand{\flm} {\varphi^l_{\mu}}
\newcommand{\fLm} {\varphi^{[L]}_{\mu}}
\newcommand{\fLma}{\varphi^{[L]}_{\mu+\varepsilon_a}}
\newcommand{\fLim}{\varphi^{[L],i}_{\mu}}
\newcommand{\z}   {E}
\newcommand{\zz}   {\varepsilon_{\z}}
\newcommand{\fIm} {\varphi^{L\rho -\zz}_{\mu}}
\newcommand{\fIim}{\varphi^{L\rho -\zz -\ve_i}_{\mu}}
\newcommand{\FL}  {\Phi^{[L]}}

\newcommand{\DL}  {D^{[L]}}
\newcommand{\DLi}  {D^{[L]}_i}
\newcommand{\fss} {\varphi^{(s,s')}_{\sigma}}
\newcommand{\fsst} {\varphi^{(s,s'),i}_{\sigma}}
\newcommand{\fssq} {\varphi^{(s,s'),i'}_{\sigma}}
\newcommand{\fssa} {\varphi^{(s,s')}_{\sigma+\varepsilon_a}}

\newcommand{\FFF}  {{\bf \sf F}}
\newcommand{\BBB}  {{\bf \sf E}}
\newcommand{\DD}  {{\cal D}}
\newcommand{\CC}  {{\cal C}}

\newcommand{\?}   {$\!$-$\!$}

\begin{document}
\pagestyle{empty}
\title{\mbox{ }\\[19.2 mm]
       SOLUTIONS OF THE KNIZHNIK - ZAMOLODCHIKOV EQUATION
       WITH RATIONAL ISOSPINS AND THE REDUCTION TO THE
       MINIMAL MODELS}

\author{ P. Furlan\\ Dipartimento  di Fisica Teorica
 dell'Universit\`{a} di Trieste, Italy\\
and Istituto Nazionale di Fisica Nucleare (INFN), Sezione
di Trieste, Italy\\
 [3.8mm] A.Ch. Ganchev{${}^*$}
\\ Theory Division, CERN, CH -- 1211~~Gen\`eve 23
\\and Istituto Nazionale di Fisica Nucleare, Sezione di Trieste
\\[3.8mm] R. Paunov{${}^*$}\\International School for Advanced
Studies (SISSA), 34014 Trieste, Italy
\\[3.8mm] and \\
\\[2.8mm] V.B. Petkova{${}^*$}\\
Istituto Nazionale di Fisica Nucleare, Sezione di Trieste
\\and International School for Advanced
Studies (SISSA), 34014 Trieste}
  \date{}
\maketitle
\begin{abstract}

In the spirit of the quantum Hamiltonian reduction  we
establish a relation between the chiral $n$-point functions,  as
well as the equations governing them,
of the  $A_1^{(1)}$ WZNW conformal theory and the corresponding
Virasoro minimal models. The WZNW correlators are described as
solutions of the Knizhnik - Zamolodchikov equations with
rational levels and isospins. The technical tool exploited are
certain relations in twisted cohomology. The results
 extend to arbitrary level $k+2 \neq 0$ and isospin values of the type
 $J=j-j'(k+2)$, $ \  2j, 2j' \in Z\!\!\!Z_+$.

\end{abstract}
       \vspace{1 cm} {\sf CERN-TH.6289/91}
\newline {\sf December 1991}
\footnotetext
{${}^*$ Permanent address: Institute for Nuclear Research and
Nuclear Energy, 1784 Sofia, Bulgaria\\ \quad \\ {}\qquad
e-mail: \quad mvxtst::furlan, ganchev@itssissa, paunov@itssissa,
petkova@itssissa}

       \vspace{-20 cm} \hspace{11 cm}

       \hspace{11 cm}
       \mbox{CERN-TH.6289/91}

       \hspace{11 cm}
       \mbox{SISSA 120/FM/91}

 \textheight =22.08cm   

\thispagestyle{empty}
\newpage
\setcounter{page}{1}
\pagestyle{plain}

\sect{Introduction}

The class of  conformal field theories introduced  by  Knizhnik
and Zamolodchikov (KZ) \cite{KZ} is expected to provide
non-perturbative solutions of the 2-dimensional WZNW Lagrangian
models at  renormalization group fixed points. The theory is
determined by its infinite dimensional symmetry described by a
pair of commuting chiral algebras -- each being a semidirect sum
of a Virasoro and a Kac \? Moody (KM) algebra $\hat g$. Because
of the holomorphic \? antiholomorphic factorization one can focus
on one of  these  chiral algebras. In the KZ theory  the
Sugawara representation of the Virasoro generators is used,  so that
the scale dimensions and the  Virasoro central charge are expressed
in terms of the level $k$ (=central charge) of the KM
algebra $\hat g$ and the weights $\Lambda$ of the representations
of the finite-dimensional subalgebra $g$. The representation of
the $L_{-1}$ generator of Virasoro by the Sugawara formula
results in the KZ differential  equation. This is a linear
first-order matrix system
of equations for the chiral $n$-point correlation functions
of the primary fields

\BL{KZ}
   \left( \frac{\dz}{\dz z_a} - {1 \over (k+h)}
   \sum_{{}^{b=1}_{b \ne a}}^n
   \frac{\Omega_{ab}}{(z_a-z_b)}\right)
  {W}^{(n)} (z_1,\Lambda_1;\dots;z_n,\Lambda_n) = 0 \, ,
\EE

$$a=1,2,....n,$$
where
\BL{}
  \Omega_{ab}= q_{\al\bt}S^{\al}_{a}S^{\bt}_{b} \,
\EE
and  $S_a^{\al}$,  $\al =1,2,\dots,$dim $g$,
acting on the $a^{\rm th}$ field,  are generators of
a representation $C_{\Lambda_a}$ of the (complex, semisimple)
algebra $g$. The matrix $q$ is the
inverse of the Killing \? Cartan form and $\,h\,$ is the dual Coxeter
number. The case most studied is the one described
by non-negative integers $\,k\,$ and a finite set of irreducible
finite-dimensional representations of $g$.

The importance of the WZNW \? KZ theories comes from the fact that
they generate a large class of rational conformal field theories.
In this paper we shall be concerned with the simplest case, the
algebra $\hat g=\hat {sl}(2,C\!\!\!\!I)$.

The idea of a hidden $\hat {sl}(2)$ symmetry in the Virasoro
minimal models (instead of the $\hat {sl}(2)\otimes \hat
{sl}(2)$ symmetry of the coset construction) originates from
\cite{KPZ}.
 If we consider  rational level and isospins,
\BL{kJ}
      J=j-(k+2)j'\equiv J_{j,j'}\  ,  \quad
      2j,2j'\in Z\! \! \!Z_+ \,  ,
\EE
\BL{rat}
   k+2=p/p'  \,,
\EE
$p,p'$ -- coprime (positive) integers, and $\,j, j'$ --
restricted furthermore to
\BL{jj'}
  1 \le 2j+1 \le p-1, \quad 1 \le 2j'+1 \le p'-1\, ,
\EE
we can rewrite the central charge and the Kac
formula for the conformal dimensions $h(j,j')$  of the minimal
models   as

\BL{c_k}
   c_k=13-6( k+2 + {1 \over k+2})\ ,
\EE

\BL{h(J)-J}
   h(j,j')\equiv h_J = \triangle_J -J =h_{\underline J}\ ,
\EE
where ${\underline J} = k+2-J-1\ $ and
 $\triangle_J$ are the Sugawara scale dimensions
\BL{sugaw}
   \triangle_J={J(J+1) \over k+2 }\ .
\EE

If instead of (\ref{rat}) we  choose $k+2=-p/p'$ we recover the
$c>25$  counterpart of the minimal $c<1$ series. More generally for
arbitrary $k+2\neq 0$ and isospins $J_{j,j'}$ as in (\ref{kJ}),
the formulae (\ref{c_k}), (\ref{h(J)-J}) parametrize the reducible
Virasoro Verma modules \cite{FFV}.

Various aspects of KM algebras at rational levels have been
studied. In \cite{KW} the class of modular representations
having characters which close under the action of the modular
group  was singled out. In the notation used here this class is
described (for $\ g=A_1\ $) by the set (\ref{kJ}), (\ref{rat}),
(\ref{jj'}), enlarged to
include the points with $2j'+1=p'$. In particular the  integer
level case $k+2=p$, $2J_a=2j_a \le k\,$,
 is incorporated with an empty ``minimal
subset'', i.e. $p'=1=2j'_a+1$.
The free field resolution of rational level
$A_1^{(1)}$ Fock modules was investigated in \cite{BF}.
The  classical Hamiltonian reduction \cite {DS} has been
carried over to the quantum case
on the level of representations of the chiral KM
algebra \cite{BO}. The Virasoro characters have
been recovered as ``residues'' of $\,A^{(1)}_1\,$
characters \cite{MP} -- an example of reduction for genus-one
zero-point functions  (see also \cite{KW2} for  higher rank
generalizations).\footnote{In \cite{MP} a different,
 parametrization and correspondence from (\ref{kJ}),
(\ref{h(J)-J}), have been used.}

The full quantum field theory is defined when the correlators
are given. Thus after understanding the reduction on the
level of the chiral algebra and representations it is natural to
study the problem in its fullest, albeit for the simplest
example of $\ g=A_1\,$. In this paper (which is an extended and
detailed version of the small note \cite{FGPP})
we address  the problem of
the  construction of the ``rational level and isospins WZNW-KZ
theory''  itself, that is the description of the correlators of
rational isospin fields on the plane as solutions of the KZ equation.
 The reduction to the corresponding Virasoro minimal theory
conformal blocks is straightforward -- at the same time, as it will
become clear, the relation to the Virasoro models
motivates the choice of the solutions.
The result holds as well for the larger set described by arbitrary
$\,k+2\ne 0\,$ and $\,J_{j,j'}\,$ as in (\ref{kJ}).

The finite matrix  KZ equation (corresponding to (half) \? integer
isospins, i.e., $j'=0$ in (\ref{kJ}))  was used in the
integer-level case to write down
 2-dimensional monodromy invariant 4-point correlation
functions  \cite{ZF}. Its simplicity
allowed one to investigate directy
 the monodromy properties of the conformal
blocks \cite{TK}, \cite{Ko} and to find explicit solutions
given by multiple contour integrals \cite{ChF}, \cite{DJMM}.
 Expressions for the correlators  based on the Wakimoto
 bosonization of the model  were
also provided \cite{BF}, \cite{Boson}.
Recently solutions were found for the case of
 an arbitrary affine KM algebra \cite{SV}
(see also \cite{Metal}).

 The integrals entering the  WZNW correlators in this
(finite matrix) case
\BL{tint}
  I^{(s)}_{\mu}(z)=
    \int_{\Gamma}\Phi_{\lambda}^{(s)}(u,z)
    \,\varphi_{\mu}^{(s)}(u,z)\ du_1\dots du_s
\EE
are of the type
first encountered in the minimal models \cite{DF}, \cite{TK2},  \cite{F}.
In (\ref{tint}) the number of  integrations $s$ is expressed
as a linear combination of the isospins $\{j_a, a=1,2,...,n\}$,
e.g., $s=j_1+j_2+\dots +j_{n-1}-j_n$, and
$\Phi_{\lambda}^{(s)}$ is a multivalued function of the
same structure as the integrand of  the thermal (i.e., labelled
by $\{ h(j,0)\}$) minimal model correlators

\BL{intPhi}
   \Phi_{\lambda}^{(s)}(u,z) = \prod_{a=1}^{n-1}
    \prod _{i=1}^s(u_i-z_a)^{\lambda_{ia}} \
    \prod_{i<j} \ (u_i-u_j)^{\lambda_{ij}}\ .
\EE

The exponents $\, \lambda_{ia}, \lambda_{ij}\ $
are specified  by the
Coulomb gas representation -- see Section 2 for a summary of the
minimal and WZNW bosonization technique. The $(n-1)$ \? vector
$\ \mu=(\mu_1, ..., \mu_{n-1})\ $ accounts for the isospin degrees
of freedom. The factor $\varphi_{\mu}^{(s)}$ is a meromorphic
function -- it can be chosen to be symmetric with respect to the
variables $\{u_i\}$, with simple poles for $u_i=z_a$
$$
  \varphi_{\mu}^{(s)}={(s-|\mu|)!\over s!}\ \left(
  \prod_{i_1=1}^{\mu_1}{1 \over u_{i_1}-z_1}\
  \prod_{i_2=\mu_1+1}^{\mu_1+\mu_2}{1 \over u_{i_2}-z_2}
  \ \dots \right.
$$
\BL{varphi}
  \qquad\qquad\dots
  \left.  \prod_{i_{n-1}=|\mu|-\mu_{n-1}+1}^{|\mu|}
  {1 \over  u_{i_{n-1}}-z_{n-1}}
  + {\rm permutations \ of \ }\{u_i\} \right)\ ,
\EE
$$
  |\mu|=\mu_1+\mu_2+ \dots \mu_{n-1}\ .
$$
In the WZNW Coulomb gas technique the meromorphic factor
(\ref{varphi})  appears with $|\mu|=s$.
It is related to a  correlator
of the $\beta$ \? $\gamma$ system necessary for the bosonization of
the KM algebra.

Viewing for a moment the insertion points $z_a$ of the primary
fields as fixed, the monodromies of
the function $\Phi_{\lambda}^{(s)}(\{u_i,z_a\})$  describe a
local coefficient system over  the space $\ X_s=\{(u_1,u_2, \dots,
u_s)\in  \ {C\!\!\!\!I}^s; u_i \not = u_j, \ u_i \not =z_a \} $;
the contour $\Gamma$ is a cycle representing an element of the
twisted homology $\,H_s(X_s, \Phi)\ $ \cite{TK2}, \cite{SV}.
Changing the integration cycle one obtains different solutions of
the KZ equation. The differential form $\varphi^{(s)}_{\mu}\, du_1
\wedge\dots\wedge du_s$ can be viewed as an element of the dual
twisted cohomology $\, H^s(X_s, \nabla_{ \Phi})\, $,
$\, \nabla_{\Phi}=\nabla-\nabla ( log \Phi )$. The integrals
$I_{\mu}$ are examples of generalized hypergeometric integrals
studied recently \cite{A},  \cite{VGZ}.

The solutions of the type (\ref{tint}) are valid  for
arbitrary $k+2 \not = 0\,$ and in particular
for rational $\ k\ $ they  describe   the ``thermal''
correlators.  We will refer in general to a spin
$J=j-j'(k+2)$ with $j'=0$ as a thermal spin while the one
with $j=0$ we will call a quasithermal spin. In both cases
the corresponding minimal model integrals are obtained using  one
type of screening charges and the factor $\Phi_{\lm}^{(s)}$ in the
thermal WZNW correlator (\ref{tint}) coincides with the integrand
of the corresponding minimal block.\footnote{In a
weaker sense we will call ``thermal'' (or ``quasithermal'') the
integrals corresponding to only one type of screening charges.
Thus, e.g., the ``thermal'' integrals (\ref{tint}) can accomodate
arbitrary $\{J_a\}$, restricted only by the condition that
$s=J_1\dots +J_{n-1}-J_n$ is a non-negative integer.}
One can expect that the generalization of
(\ref{tint}) for arbitrary rational spins will involve the
general non-thermal integrand. The problem is to generalize the
factors $\varphi_{\mu}$. We recall that in  the $A_1^{(1)}$
Coulomb gas there are two solutions for the screening currents.
In the description of the free field resolution \cite{BF} one can do
with only one type of screening charges. On the other hand, to
define a general correlator with rational spins (\ref{kJ})
one is forced to consider both types of screening charges. This
lies at the heart of the problem since the second charge involves
a  non-integer power ($-(k+2)$) of the bosonic field $\beta$. If one
defines \cite{D} the correlator of rational spin fields by
analytic continuation in $J_a$ and $k$ of the bosonization
formulae, valid  for integer $2J_a$ and $k$, the factor
$\varphi$ -- related to an expectation value of the
$\beta$ \? $\gamma$ system -- is no longer a meromorphic function.
Thus effectively one has changed the local coefficient system
defined by $\Phi$. This may change the fusion rules if compared
with the fusion rules of the corresponding minimal model,
irrespectively of the values of $p,p'\,$, as happens in the
simple example investigated in \cite{D}.
Hence the direct connection between the correlators of the two
types of models is eventually lost.

The  difficulty is related to the fact that we have to deal with
infinite dimensional representations of the isospin  algebra
$\,sl_I(2,C\!\!\!\!I)\,$ when we allow non-integer $2J$. Putting the
isospin and the coordinate $\,sl(2,C\!\!\!\!I)\,$ symmetry
on equal footing one can realize the generators $\,S^0,S^+,S^-\ $ of
$\ sl_I(2,C\!\!\!\!I)\ $ as differential
operators with respect to a second complex variable $x$, i.e.,

\BL{hl}
   S^0=2x \dz_x -2J,\quad  S^-=-\partial_x,
   \quad  S^+=x^2\dz_x-2Jx \,.
\EE
If in these differential operators we replace $(x,J)$ by
$(z, - \triangle)$ we obtain the generators  $\ \- 2L_0$,
$-L_{-1}$, $L_1\ $  of the $\,sl(2,C\!\!\!\!I)\,$ subalgebra of
Virasoro.
This realization has been used in \cite{ZF} in the
(half)-integer isospin case $J=j$ where there exists an invariant
finite dimensional subspace of fields $\Psi^J(x,z)$, spanned by
all polynomials of $x$ of highest degree  $2J$.
(In a Minkowski space framework the variables $x$ and $z$
are real and $S^{\alpha}$ (or $L_{\alpha}$), $\alpha = \pm 1, 0$,
represent the generators of an induced representation of the
group $SL(2,I\!\! R)$ labelled by $J$ (or $-\triangle$).)
The (left) action of the generators $X_n^{\alpha}$ of the KM algebra
$\hat g$ on the primary fields $\Psi^J(x,z)$ reduces to
$S^{\alpha}$, as defined in (\ref{hl}), times $z^n$. The field
$\Psi^J(x,z)$ can be treated as a highest (or lowest) weight
state with respect to a right action of $\hat g$, thus generating
a module of descendants. (See \cite{ZF} for explicit formulae which
carry over to arbitrary values of $J$.)

Inserting the expression (\ref{hl}) for the generators   in the
KZ system (\ref{KZ}) we can look for solutions of the
resulting equations in the space of all $n$-point invariants
with respect to  both $\,sl(2,C\!\!\!\!I)\,$ algebras  for
arbitrary values of the isospins. The starting point of our work has
been the following simple observation. The 2- and
3-point chiral functions are given by the factorized expressions
\BL{}
   W^{(2)}(x_1,z_1,J;x_2,z_2,J)=  (x_{12})^{2J} \
   (z_{12})^{-2\Delta_{J}},
\EE

\BL{}
   W^{(3)}(x_1,z_1,J_1;...;x_3,z_3,J_3;)= C_{J_1 J_2 J_3}\,\,
   { x_{12}^{J_1+J_2-J_3}\ x_{23}^{J_2+J_3-J_1}\
     x_{31}^{J_1+J_3-J_2} \over
   z_{12}^{\Delta_{J_1}+\Delta_{J_2}-\Delta_{J_3}}\
   z_{23}^{\Delta_{J_2}+\Delta_{J_3}-\Delta_{J_1}}\
   z_{31}^{\Delta_{J_1}+\Delta_{J_3}-\Delta_{J_2}}\ } .
\EE

Setting $x_a=z_a$ and using (\ref{h(J)-J}) we immediately obtain
the respective functions of the minimal models. This simple
factorization is no longer valid for higher $n$-point functions.
Instead we can try to find solutions for the correlators in
terms of power series of $(x_a-z_a)$. For example,
for the 4-point function we can write, using the projective invariance,
\BL{x-z44}
  W^{(4)}(\{x_a,z_a\}) = {\rm prefactor}  \sum_{t=0}
  (\ux-\uz)^t\  C_t(\uz) \ ,
\EE
where $\ux$ and  $\uz$ are  anharmonic ratios.

The KZ equation for the 4-point function  translates
into an (infinite in general) matrix  differential equation for the
coefficient functions $C_t$ in (\ref{x-z44}).
It is a generalization of the finite
KZ \? ZF system (for Knizhnik, Zamolodchikov, Fateev)
written down in the integer level case.
It is instructive to start first with the chiral solutions of
the finite KZ \? ZF system.  In Section  3
we describe the set of integrals  serving
as coefficients in the $(x-z)$ \? expansion of the $n$-point chiral
correlators in the thermal case.
In particular for $n=4$
these integrals  $\ \{C_t(\uz)= B_t \ I_t^{(s)}(\uz)\}\ $
 are,  up to  the numerical coefficients $\ B_t \,$,
of the type in (\ref{tint}), (\ref{intPhi}),  (\ref{varphi}),
with $\ \mu=(0,t,0), \ t=0,1,...,s\ $.
These solutions are equivalent to the
representations of the $n$-point functions coming from
bosonization,  \cite{BF}, \cite{Boson},
 or other related representations
studied in the literature \cite{DJMM}.
The latter can be written as $x$-expansions and they
involve $\varphi^{(s)}_{\mu}$ with $\ |\mu|=s\ $. The tool used
to prove the equivalence of the various integral representations
in the thermal case is a set of
linear relations, derived by Aomoto \cite{A} in the study of the
generalized hypergeometric integrals and the twisted cohomology groups
related to them. The $(x-z)$ \? expansion can be
interpreted as providing a new basis in the top (i.e., the
$s^{\rm th}$) twisted cohomology which includes
$\varphi_0^{(s)}\simeq 1$ as an element or --  upon integration -- the
integral $I_0^{(s)}$, which coincides with the corresponding
minimal model integral. This basis is also suitable for the reduction
of the KZ \? ZF system studied in Section 3.2. Indeed the integrals
in the $(x-z)$ \?
expansion (\ref{x-z44}) can be written as
\BL{C=LC}
   \gamma_t\,  I^{(s)}_t(z)=
  {\cal L}_t (z,\partial)\, I^{(s)}_0( z)
  \,,\qquad t=1,2,\dots
\EE
with ${\cal L}_t$ a differential operator of order $t$ and
$\,\gamma_t\,$ -- a numerical coefficient (see
formulae (\ref{BPZ}), (\ref{bpzn})). Since the KZ \? ZF in
the thermal case  is a finite system, for a certain $t$ the
left-hand side of (\ref{C=LC}) vanishes and we obtain the
corresponding BPZ
differential equation for the minimal model correlators describing
the decoupling of the Virasoro singular vectors \cite{BPZ}. Thus in
the thermal case we have two theories that are independently
well defined  -- the WZNW having correlators that are solutions
of a finite KZ \? ZF system and
the corresponding minimal model. Taking $x\to z$ we get a
reduction of the first to the second. Vice versa, given the
minimal blocks we can recover the WZNW correlations
according to  (\ref{C=LC}). Note that the situation in  the
integer level case, which does not have a minimal model
analogue, is somewhat different. Depending on the specific
 combinations of the spins, the series
(\ref{x-z44}) appears to be effectively truncated from below \cite{ZF}
so that the limit $\ x_a \to z_a\ $ is trivial.

In the non-thermal case  the
WZNW conformal theory is not uniquely defined.
The KZ \? ZF system is infinite,
reflecting the fact that for non-thermal
spins $J$ the representations of $\,g\,$ are infinite-dimensional.
In Section 2.2 we briefly discuss the various solutions coming from
the analytic continuation of the
bosonization technique. Motivated by the consideration in the
thermal case  we generalize in Section 4.1 the $(x-z)$
 \? expansion finding explicit solutions for the coefficients.
Now the two types of solutions are no longer equivalent.
While the former involves, as mentioned, multivalued functions
  $\ \varphi_{\mu}$, $\, |\mu|=s-(k+2)s'\, ,$ the latter is
described by ``meromorphic'' factors $\ \varphi_{\mu}^{(s,s')}$,
$\ \mu_a \in Z\!\!\!Z_+\ $,
generalizing (\ref{varphi}).
 The hard part is to find these
non-thermal meromorphic factors
(see formulae (\ref{nphi}), (\ref{Lphi})). The corresponding
integrals satisfy a system of linear and recursion relations
which have a cohomological interpretation and
generalize the ones in the thermal case. All this highly
technical part is put in a separate section -- Section  5.
The solutions of the KZ \? ZF infinite system  described  in
Section  4.1  are such that for generic spins they
reduce in the limit $ x \to z$ to the minimal correlations. One
can look at the resulting WZNW theory, described by an infinite
set of integrals $\{I_{\mu}^{(s,s')}, \ |\mu|=0,1,\dots\}$, as being
``generated''  uniquely by the corresponding minimal one
$\,I_0 ^{(s,s')}$. The correlators in this theory have the same
braiding properties, under simultaneous change of $x$ and $z$, and
hence the same fusion rules as the corresponding minimal Dotsenko \?
Fateev (DF) blocks. These ``meromorphic'' solutions can be viewed as
analogues of the ``odd'' combinations of KM characters in \cite{MP}
(labelled by $J$ and $\uJ$)
which transform with the minimal model modular matrix.

The rationality of the factor $\ \varphi_{\mu}^{(s,s')}\ $, also
ensures  the actual truncation  of the infinite KZ \? ZF
matrix  system of equations to a finite system, which
can furthermore be reduced to a BPZ type equation for the
minimal integrals. The origin  of this truncation is the
existence of additional linear relations for the integrals
which reflects the fact that any rational form can be expressed
up to an exact form in terms of a finite basis in the
relevant twisted cohomology \cite{A}. Section 4.3 contains a preliminary
discussion of this point which deserves  further study.
In Section  4.2 and in the Appendix we describe,
using the example of $\,n=4\,$, an explicit algorithm to
obtain these relations in the quasithermal case. It is
based on a duality property of the WZNW models with $k+2$,
$\,\{J_a\}\,$ and with $\hat k +2=1/(k+2)$, $\{\hat J_a=-J_a/(k+2)\}$
-- both generated by the same minimal correlation functions.
On the other hand the additional relations, and hence the
truncation, can be obtained equivalently taking into account one of the
algebraic null vector decoupling
 equations, originating from  singular vectors of the
 $\ A_1^{(1)}\ $  Verma modules.
This is discussed and illustrated in examples in
Section  4.3. Section 6 contains a  discussion of the open problems.

\renewcommand{\theequation} {\thesubsection.\arabic{equation}}
\sect{Free field realizations of the   minimal and $A_1^{(1)}$ - models}
\subsection{Minimal models}

The free field  realization of the minimal models \cite{DF}
utilizes a free bosonic chiral field $\ff(z)$ with a logarithmic
2-point function
and a modified energy momentum-tensor reproducing the
central charge (\ref{c_k}). A primary field
$\Psi^{h(j,j')}(z)$ of conformal dimension $h(j,j')$ given by
(\ref{h(J)-J}) is represented by a vertex operator
\BL{V}
   V_{\al}(z) = :e^{i\al\ff(z)}:
\EE
of charge
\BE
  \al=\al(j,j')=  -j\al_- - j'\al_+=-J\al_- \equiv \al_J \ ,
  \qquad  {\rm or} \qquad
  \al=2\al_0- \al(j,j')= \al_{\uJ} \,  ,
\EE
where
\BE
2\al_0=\al_+ +\al_-  \, , \quad
\al_{\pm}=\pm (k+2)^{\pm1/2}\, .
\EE

The vertex operators in a correlator must satisfy a charge
conservation condition shifted by the charge $\ -2\alpha_0\ $ at
infinity.  In order to be able to write non-trivial
correlators one needs  (non-local) fields of non-zero charge and
zero conformal dimension so that the insertion of such fields
inside correlators could fit the charge balance without changing
the conformal properties. These are the screening charges --
contour integrals of the conformal dimension 1 screening currents
\BL{V+-}
   V_{\pm}(z) =  V_{\al_{\pm}}(z) \,.
\EE

The correlators of $n$ primary fields
$\Psi^{h_a}(z_a)$, $\ h_a=h(j_a,j'_a), \  a=1,\dots, n$
are thus represented by a multiple contour integral
\BA{calG}
   {\cal S}_{\Gamma}^{(n)}(z_1,\dots,
   \!\!\!\!\!\!\!\!\!\! && z_n)    =
   {\langle}0|\Psi^{h_1}(z_1) \Psi^{h_2}(z_2) \dots
   \Psi^{h_n}(z_n)|0{\rangle}_{\Gamma} =
\\ \nonumber
   && \int_{\Gamma} du_1\dots du_s \,  dv_1\dots dv_{s'}{\langle}
   V_{\al_1}(z_1)
   \dots V_{\al_n}(z_{n})
   V_-(u_1)\dots V_-(u_s)V_+(v_1)\dots V_+(v_{s'})
  {\rangle}_{\al_0} \, .
\EEA
Without lack of generality we can choose
$\al_a=\al(j_a,j'_a)$, $a=1,2,...,n-1$ and
$\al_{n}=2\al_0-\al(j_{n},j'_{n})$
so that charge neutrality implies
\BL{ss'}
   s=j_1+j_2+\dots+j_{n-1}-j_n \, ,\qquad
   s'=j'_1+j'_2+\dots+j'_{n-1}-j'_n \, .
\EE
Using the projective invariance of the conformal blocks,
(\ref{calG}) can be written as
\BL{G=fI}
  {\cal S}^{(n)}_{\Gamma}(z_1,\dots,z_n)
  =f(\{z_a; h_a\}_{a=1}^n)
  \, \prod_{1\le a<b\le n-1}(\uz_{a}-\uz_b)^{2J_aJ_b/(k+2)}
  \ I_{\Gamma}^{(s,s')}(\uz)\ ,
\EE
where

\BL{zeta}
  \uz_a = \frac {z_{1a} \, z_{n-1,n}}{z_{1,n-1}\, z_{an}},
\EE
for $\  a=2,\dots,n-2\ ,$ are the anharmonic ratios (the
moduli  of the $n$-punctured sphere) and the prefactor in
(\ref{G=fI}) is

\BL{pref}
  f(\{z_a; h_a\}_{a=1}^n)=
    \prod_{a=1}^{n-1} \ (z_a-z_n)^{-2h_a} \
  \left(\frac {(z_1-z_n)(z_{n-1}-z_n)}
  {(z_1-z_{n-1})}\right)^{\sum_{a=1}^{n-1} h_a-h_n} \,.
\EE

The data in (\ref{calG}) or (\ref{G=fI}) consist of the
integration contours $\Gamma$ and the spins $J_a$ attached to
the punctures $z_a$ (or $\uz_a$). For short,
we will denote the respective $(n-1)$ \? vectors by $z=(z_a)$ (or
$\uz=(\uz_a)$) and $\ {J}=(J_a)$, $a=1,\dots,n-1$.
Instead of $J_n$ we will specify $(s,s')$ or equivalently
$S=s-s'\al_+^2$. Most of the time we will skip indicating part
of the data when there will be no danger of confusion.

The computation of the  integrand  in (\ref{calG}) gives
\BL{I}
   I_{\Gamma}^{(s,s')}(\uz)= \int_{\Gamma}
   du_1\dots du_s\   dv_1\dots dv_{s'}\
  \Phi^{(s,s';\al_-^2)}_{{J}}(u_i,v_{i'};\uz_a)
\EE
where the multivalued function $\Phi$ is given by (cf.
(\ref{intPhi}))
\BL{PhiPhi}
   \Phi^{(s,s';\nu)}_{{J}}(u_i,v_{i'};z_a) =
   \Phi^{(s;\nu)}_{{J}}(u_i;z_a)\,
   \Phi^{(s';1/\nu)}_{-\nu {J}}(v_{i'};z_a)
   \prod_{i=1}^s\prod_{i'=1}^{s'} {1\over (u_i-v_{i'})^2}\, ,
\EE
and
\BL{Phi}
   \Phi^{(s;\nu)}_{{J}}(u_i;z_a) = \prod_{i<j}^s (u_i-u_j)^{2\nu}
   \prod_{i=1}^s\prod_{a=1}^{n-1} (u_i-z_a)^{-2J_a\nu} \, .
\EE
Recall that we denote $J_a=j_a - j'_a \al^2_+$, thus for
$\nu=\al_-^2$ we have $-\nu J_a=j'_a - j_a \al_-^2$.

Making different choices for the integration  contours $\Gamma$
we can obtain a basis in the space of $n$-point functions.  One
possibility to get a ``basic'' set is to use the notion of a
chiral vertex operator
${}_{h_1}\Psi^{h}_{\,\,h_2}(z) = P_{h_1}\Psi^{h}(z) P_{h_2}\,$,
where  $P_{h}$ is a projector on the Virasoro representation
of highest weight $h$. These operators can be realized
as screened vertex operators  \cite{F} with screening charges
integrated along a double loop
(see also \cite{FS}). Inserting projectors between primary fields we obtain an
$n$-point conformal block with specified spins in the
intermediate channels labelled by a vector ${K}=(K_a)$ (with
$K_a=k_a-k'_a\al_+^2$) such that $K_1=J_1$, $K_{n-1}=J_n$.
The corresponding contour $\Gamma_{{K}}$ is a configuration of
nested double loops
( i.e. the $i^{\rm th}$ contour consists of a loop around $\ z_{i+1}\ $
and a loop encircling all $\ z_1,\ z_2,\  \ldots ,\  z_{i}\ $).
We will call
``admissible'' the set of such contours when the intermediate
channels $(K_a)$ are consistent with the fusion rules,
 i.e., for every triple
$(K_{a-1},J_a,K_a)$, $a=2,\dots,n-1$ the fusion rules are
satisfied. Another such choice is provided by a subset of
the contours connecting the singular points \cite{DF}.
The minimal models fusion rules multiplicities
$N^h_{h_2 h_1} = ( N^h )_{h_2 h_1}$  (= the number of non-zero
chiral vertex operators ${}_{h_1}\Psi^h_{\,\,h_2}$) have the
factorized form
\BL{N=NN}
   N^{h(j,j')}_{h(k_2,k'_2)\,h(k_1,k'_1)} = N^j_{k_2\,k_1}\
   N^{j'}_{k'_2\,  k'_1}
\EE
where $N^j_{k_2\,k_1}$ is symmetric with respect to all indices
and $N^j_{k_2\,k_1}=1$ if and only if

\BL{2.2.21}
| k_1-k_2| \leq j \leq \min (k_1+ k_2, p-2-k_1-k_2)\, ,
\EE
otherwise $N^j_{k_2\,k_1}=0$, and analogously for the primed
spins. For fixed $J_1,\dots,J_n$  the dimension of the space of
$n$--point correlators with
$\Gamma_{K}$  is thus
\BL{N}
   \left(N^{h(J_{n-1})}\dots N^{h(J_3)}N^{h(J_2)}
   \right)_{h(J_n)\, h(J_1)}\ .
\EE

This bears some features in common with the set of
$sl(2,C\!\!\!\!I) $ - invariant tensors. Actually the relevant
structure is related to a deformation of this algebra, the
quantum universal enveloping
algebra $U_q(sl(2,C\!\!\!\!I))$ when the
deformation parameter is $q=exp ( 2\pi i\, p/p')$ or
$q'=exp (2\pi i\, p'/p)$ (see e.g. \cite{GS} and
references therein).

The formalization of the above description is in terms of twisted
homology, i.e., homology on the space $X_{s+s'}$ of the positions
$u_i$, $i=1,\dots,s$ and $v_{i'}$, $i'=1,\dots,s'$ of the
screening charges (with no two of the points $z_a$, $u_i$, $v_{i'}$
coinciding) with coefficients in the local system defined by the
multivalued function $\Phi$. The general statement is that the
integration contour $\Gamma$ is a cycle in the top twisted
homology and, as $\Gamma$ runs over a basis in homology, the
integrals $\int_{\Gamma}\Phi$ give a basis in the space of
correlators.

These statements actually need a lot of refining in the case of the
generalized hypergeometric integrals (\ref{I}) which appear in the
minimal theories. In fact there are quite a few points that still
await a precise mathematical treatment. These  questions
do not constitute the core
of our discussion and we will be content with several remarks.

The appropriate homology depends on the type of contours.
When the exponents  in  (\ref{PhiPhi}) are rational numbers,
the usual assumptions of ``non-resonance'' of these exponents \cite{A},
\cite{VGZ}
are not valid in general and the dimension of the
``homology of double loops'' can be different from the
dimension of the respective homology of
contours which turn on singular points. Accordingly,
depending on the choice of the contours, the vertex representation
provides the conformal blocks of (at least) two
essentially different types of
models. The minimal one -- in which the symmetry $\ \alpha_J
\rightarrow 2\alpha_0 -\alpha_J \ $ is preserved -- is described by
an admissible set of contours. The second leads to ``unphysical''
(i.e.,  not in the domain (\ref{jj'})  contributions and it admits an
interpretation in terms of quantum group modules.
This  is also related to the peculiarities of the quantum group
representation theory when the parameter $q$ is a root of unity.

Finally let us mention one more point that deserves a
rigorous treatment. The general non-thermal multivalued function
(\ref{PhiPhi}) involves integer exponents and
has a peculiar factorised form suggesting
that  we should consider a factor of homology -- identifying
cycles that differ by interchanging $u$ and $v$ contours.
Thus this factor homology factorizes into $u$ cycles times
$v$ cycles (in agreement with the results of \cite{DF}) and this leads
to the factorization of the fusion rules
(\ref{N=NN}).

\subsect{$A_1^{(1)}$ -- models}

The free field realization  of these models involves
the  free field $\ff(z)$ and a pair
of generalized free  bosonic fields
$(\bt, \ga)$  with a 2--point function

\BL{bg2}
  {\langle}\bt(z_1)\ga(z_2){\rangle} = \frac 1{z_1-z_2} \,.
\EE
The chiral algebra is  expressed in terms of
this triple of bosonic  fields.

Let all  $2J=2j$ be non-negative  integers. The primary fields
$\  \Psi^J(x,z)\ $ are then  given by the polynomials

\BL{Psi(x)}
  \Psi^j(x,z) = \sum_{m=-j}^{j} {2j\choose j-m}
  \, x^{j+m}\,  \Psi^j_m(z)\, ,
\EE
where the coefficients   $\Psi^j_m(z)$  can be  represented  as

\BL{2.3.4}
  \ga^{j-m}(z)\  V_{\al_j}(z)
\EE
using the vertex operator (\ref{V}) of charge
$\ \al_j=-j\alpha_-$. The screening current is given by
\BL{2.3.6}
  \Psi_-(u)=\bt(u)V_-(u)
\EE
with  $V_-(u)$ defined in (\ref{V+-}).

Another representation for the operator  $\Psi^j_m(z)$ is
provided using the vertex operator of conjugate charge $2\al_0
-\al_j$. It involves also the second  bosonic field $\bt$ and
it looks simple only for the state with  $m=j$. Namely, within
the correlation functions we can represent  $\Psi_j^j(z) $ by
\BL{2.3.5}
   \bt^{2j}(z)\  V_{2\al_0 -\al_j}(z)\, .
\EE
Exploiting the isospin projective invariance of the correlators
we can always put one of the fields at $x=\infty$, so that only
the component with $m=j$ survives in  $x^{-2j}\Psi^{j}(x,z)\ $
in this limit.

We stress that the representation (\ref{2.3.5}) is consistent
only for the field averages. They are computed  using the
factorization of the $\ \varphi$ \? and
$\ \bt,\ga$ \? dependent contributions; see, e.g., \cite{D} for a
detailed description  of the bosonization in the operator
framework.   Thus the correlator of $n$ fields
$\Psi_{m_a}^{j_a}(z_a)$, $a=1,\dots,n$ is expressed  by the
average of an equal number  of $\bt$ and $\ga$ fields, times a
vertex correlator in the presence of a background charge
$-2\al_0$ of the type encountered in the minimal theory, namely

$$
   {\langle}0|\Psi_{m_1}^{j_1}(z_1)  \dots
  \Psi_{j_n}^{j_n}(z_n)|0{\rangle}_{\Gamma}=
  \int_{\Gamma} du_1\dots du_s
   {\langle}\ga^{\tau_1}(z_1)\dots
 \ga^{\tau_{n-1}}(z_{n-1})\bt^{2j_n}(z_n)\bt(u_1)\dots\bt(u_s)
 {\rangle}
$$

\BL{x}
  \qquad\qquad \cdot {\langle} V_{\al_{j_1}}(z_1)
  \dots V_{\al_{j_{n-1}}}(z_{n-1}) V_{2\al_0-\al_{j_n}}(z_{n})
   V_-(u_1)\dots V_-(u_s)  {\rangle}_{\al_0}\ ,
\EE
where

\BL{tau}
  \tau_a = j_a - m_a, \ a=1,\dots,n-1\, ;\
  |\tau|=\tau_1+\tau_2+\dots + \tau_{n-1}\, , \quad
|\tau|=2j_n+s \ ,
\EE
and as above  the charge conservation condition
implies that $ s=j_1+j_2+\dots+j_{n-1}-j_n \, \ $,  so that
$m_1+m_2+...+m_{n-1}+j_n=0$.

Using the projective invariance we can write the $n$-point
function of the fields $\  \Psi^J(x,z)\ $ as

\BL{cor}
   {W}^{(n)}(\{x_a,z_a,J_a\}_{a=1}^n) =
   f(\{ x_a, -J_a\}_{a=1}^n)\,f(\{ z_a, \triangle_a\}_{a=1}^n)
   \,(2J_n)!
   \prod_{1 \le a < b \le n-1} \uz_{ab}^{\,2J_aJ_b/(k+2)}
  \  G^{(n)}_{J}(\ux,\uz)\ ,
\EE
where $f$ is the prefactor defined in (\ref{pref}).
Most of the time we will work with correlators in
which the last point is sent to infinity
\BL{W_inft}
   {U}^{(n)}(x,z) \equiv
   {1\over (2J_n)!}\, \lim_{{}^{x_n\to\infty}
     _{z_n\to\infty}} x_n^{-2J_n}\,z_n^{2\triangle_n}\,\,
    {W}^{(n)}(\{x_a,z_a,J_a\}_{a=1}^n)
    =\prod_{1\le b<a\le n-1}z_{ab}^{2J_aJ_b/(k+2)}
    \,\, G^{(n)}_{J}(x,z)\, .
\EE

For thermal
isospins $\ J_a=j_a\ $ (\ref{Psi(x)}) and (\ref{x}) give
\BL{G}
   G^{(n)}_{j}(x,z)=  \sum_{|\tau|=s+2j_n}\, \prod_{a=1}^{n-1}
  {2j_a \choose \tau_a}   x_a^{2j_a-\tau_a}
  \,\,{\cal G}_{(\tau_1,\dots,\tau_{n-1})}(z) \ ,
\EE
\BL{cal G}
   {\cal G}_{(\tau_1,\dots,\tau_{n-1})}(z) =
   \int_{\Gamma}du_1...du_s\
   \,\,\Phi^{(s;\nu)}_{J}(u_i;z_a)\
   \,\,\phi_{\{\tau_1, ..., \tau_{n-1}\}}(u_i;z_a)\, .
\EE

The correlator $\ \phi_{\tau}\ $ of the generalized free fields
$\ \bt,\ga\ $ is computed by Wick contractions using the 2-point
kernel  (\ref{bg2}). Here we shall write down  the explicit
expression for a more general  bosonic correlator with arbitrary
non-negative integer  powers of the $\ \bt, \ga\ $ fields, which
will be useful in what follows. To do that let us first  introduce
some notation.

The indices $a,b(=1,\dots,n-1)$ will be reserved
for the positions of the vertex operators $z_a,\, z_b$ and the
indices $i,j(=1,\dots,s)$ for the coordinates $u_i,\, u_j$ of
the screening currents. The convention will be, if not
indicated otherwise, that  sums and
products in $a$ or $b$ will run over $1,\dots,n-1$, while
those in $i$ or $j$ -- over  $1,\dots,s$.
 For an $(n-1)$ \? vector
of non-negative integers we will use the notation $\mu=(\mu_a)$
(also $\tau$ and $\sigma$) and the sum of its elements
$|\mu|=\sum_a\mu_a$.

 Let $\mu=(\mu_a)$ and $l=(l_i)$ be    $(n-1)$ \?$\,\,$ and $s$ \? vectors
of non-negative integers and such that $\ |\mu| =|l|$.
Consider a rectangular matrix $A=(A_{ia})$ with integer valued
entries $A_{ia}=0,1,2,...$ and such that
\BL{l}
  \sum_{a=1}^{n-1} A_{ia} = l_i
\EE
and
\BL{sA=ma}
  \sum_{i=1}^{s} A_{ia} = \mu_a \, .
\EE
Denote
$$
  \ff_A^l(u,z)=\frac{\prod_a\mu_a!}{\prod_{i,a}A_{ia}!
(u_i-z_a)^{A_{ia}}}
$$
and
\BL{ff}
  \ff_{\mu}^l(u,z)=\sum_A\ \ff^l_A(u,z)
=\sum_A \ \frac{
\prod_a\mu_a!}{\prod_{i,a}A_{ia}!\ (u_i-z_a)^{A_{ia}}} \, ,\quad
|\mu| =|l|\, ,
\EE
where the sum is over all $\{ A_{ia}\} $ satisfying (\ref{l}), (\ref{sA=ma}).

Let   $\tau=(\tau_1, ..., \tau_{n-1})\  $ be  a  $(n-1)$ \? vector
with non-negative integer entries. Using the above notation,
the bosonic correlator

\BL{phi(l)}
\phi_{\tau}^{l}(u_i;z_a)={1 \over \prod_{i=1}^{s}l_i! (2j_n)!}
\lim_{z_n\to\infty} (z_n)^{2j_n}
{\langle}0|\ga^{\tau_1}(z_1)\dots
\ga^{\tau_{n-1}}(z_{n-1})\bt^{2j_n}(z_n)\bt^{l_1}(u_1)\dots\bt^{l_s}(u_s)
|0{\rangle}\,
\EE
can be written as
\BL{phil}
  \phi_{\tau}^l =  \sum_{\mu:\,\,|\mu|=|l|} \prod_a
  {\tau_a\choose\ma}\ \ff_{\mu}^l \, .
\EE

In the thermal case $\ l=\rho \equiv (1,1,...,1)\ $, so that $|l|=s$
and $\ A_{ia}=0,1\ $.
 Hence
the integrand in (\ref{cal G}) is a modification with a meromorphic
factor of the minimal thermal multivalued function (\ref{Phi}).
 Multiplying (\ref{phil}) (for $l=\rho\ $) by the
contribution of the vertex operators and integrating, we get
for the  thermal conformal blocks (\ref{cal G})
\BL{G2}
  {\cal G}_{(\tau_1,\dots,\tau_{n-1})}(z) =  \
  \sum_{\mu : |\mu|=s}\,\,\prod_a{\tau_a\choose\ma}\,
  I^{(s)}_{\mu\,\,\Gamma}\,(z) ,
\EE

\BL{}
  I^{(s)}_{\mu\,\,\Gamma}=\int_{\Gamma} du_1\dots du_s \,\,
 \Phi_{j}^{(s;\nu)}(u,z)\, \,   \ff_{\mu}(u,z).
\EE

Note that in that case  $\ \ff_{\mu} (\equiv
\ff_{\mu}^{\rho}) \ $ itself can be
written as a $(\bt, \ga)$ correlator

\BL{db}
 \ff_{\mu}(\{u_i,z_a\})=(-1)^s
\langle \gamma(u_1)... \gamma(u_s)
\beta^{\mu_1}(z_1)....\beta^{\mu_{n-1}}(z_{n-1})\rangle\, ,
\EE
which coincides
 with
$\ \varphi_{\mu}^{(s)}\ $ in  (\ref{varphi}),
taken  for $\ |\mu|=s\ $.

The integral representation (\ref{G2}) can be extended to
rational  isospins $J_a$ if $s'=j'_1+j'_2+\dots +j'_{n-1}-j'_n=0$,
so that $J_1+J_2+\dots+J_{n-1}-J_n=s$. However,
to describe the correlators for  arbitrary
rational spins $J_a$ one needs  a second screening
charge operator, which is  provided by

\BL{2.3.8}
  \Psi_+(v)=\bt^K(v)V_+(v)\, ;\quad  K=-(k+2)\, .
\EE
Now
the natural
idea \cite{D} is to compute the correlations using  the two types of
screening currents and
 the vertex operators
(\ref{2.3.4}), (\ref{2.3.5}) for general isospins, i.e.,
replacing $j_a\ $ by $J_a\ $ and  assuming first  that all
$J_a$ and $K=-(k+2)$ are integers. Then  we have to continue in
some way the final result.
The equality of the number of   $\bt$ and $\gamma$ fields
implies the  condition
\BL{taun}
|\tau|=2J_n+S, \ S=s+s'K; \ \ s^{(')} = j^{(')}_1+j^{(')}_2+
\dots j^{(')}_{n-1}-j^{(')}_{n}\ .
\EE
The correlator of the vertex operators gives the same
multivalued function (\ref{PhiPhi}) as in the general
minimal model case, while the bosonic correlator (\ref{phil}) is
replaced by

\BL{phin}
  \phi_{\tau}^{(\rho,\rho K)}(u,v, z)=\sum_{|\mu|=S}
  \prod_a {\tau_a \choose \ma} \sum_{|\alpha|=s } \prod_a
  {\mu_a  \choose \alpha_a}
  \varphi_{\alpha}^{\rho}(u,z)\varphi_{\mu-\alpha}^{\rho K}
  (v,z) \, .
\EE

The problem now is that this  correlator will not be a
meromorphic function any more when we continue back to the true
rational values of $J_a$ and $K$. In particular in the
quasithermal case,  i.e., $\ s=0, S=s' K\ $, involving only the
second  screening charge operator (\ref{2.3.8}),
we can continue  (\ref{ff}) for non-integer $K  \ $ and
$\mu$ such that, $\ |\mu|=S=s'K\ $,
choosing the  $\ n-2\ $  independent  entries of the
vector $\ \mu ,\ |\mu|=S, \  $ and
$\ (n-2)(s'-1)\ $ of the parameters
  $A_{ia},\ $  to be non-negative integers.
The rest of the parameters  are expressed using the restrictions
in (\ref{l}),
(\ref{sA=ma}) with  $\ l_i=K \ $ and thus we end up with
infinite series, representing a non-meromorphic  factor
$\ff_{\mu}^{\rho K}$.
 More generally  we can choose a matrix $A_0$ satisfying
(\ref{l}), (\ref{sA=ma}), and take the sum in (\ref{ff})
over all matrices $A$ satisfying these restrictions and differing
from
$A_0$ by integer matrices.
Similarly we can generalize the
$x_a$ \? expansion  (\ref{x})
choosing an arbitrary vector $\  \tau^{(0)}\ $
satisfying $\ |\tau^{(0)}|=2J_n+S\ $. Then we can sum over
$\ \tau =(2J_1,
\tau_2,...\tau_{n-1}), $
 satisfying this condition, which differ from $\
\tau^{(0)}\ $ by  integers.
(Recall that $\ux_1=\uz_1=0$,
 so that $\tau_1=J_1-M_1=2J_1$. All
factorials are replaced by  $\Gamma$ - functions.) This gives
meaning to the infinite series,
which replace the finite sums in (\ref{G}).

In general   these
extensions,  leading to different non-meromorphic factors
$\ \ff_{\mu}^{\rho K}\ $,  are  inequivalent.

 For example for $\ n=4\ $ and $2J_2=2J_3=K, J_1=J_4,$ i.e.,
$\
s'=1\ , S=K, \ $ (the case considered in
\cite{D}) we can choose
$\ \tau^{(0)}=(2J_1,K,0)\ $
 and $\ \mu=(t,K-t,0)\, , t=0,1,... $ -- then the factor (\ref{phil})
 is simply \footnote{Here and in what follows we  keep for simplicity
the notation $a!$ instead of $\Gamma(a+1)$. }
\BL{s'=1}
  \phi_{\tau}^{K\rho}(\uz)=
  \sum_{t=0}^{\infty} \ {2J_1 \choose t} {K \choose t}
  \frac{K!}{ u^{t}(u-\uz)^{K-t}} \, .
\EE
Summing furthermore in (\ref{G}) over
$\tau=(2J_1,K-l,l)\  , l\in  Z\! \! \! Z_+ \ $
we get a power series in $\ \ux \ $ for the
correlator. This corresponds to $\ M_1=-J_1 \ ,
\ M_2=-K+l \ , \ M_3=K-l \ , \ M_4=J_1 \ , \
l\in Z \! \! \! Z_+ \ $, i. e. to the choice of
lowest weight representations
of $sl(2)$ sitting at the first and the second points and highest
weight representations at the last two points.
Another choice, made in \cite{D}, is provided by
$\tau^{(0)}=(2J_1,0,K)\ $,   $\ \mu=(t,0,K-t)\, ,
t=0,1,..., $ so that in (\ref{s'=1}) $(u-\uz)$ is replaced by $(u-1)$.
This choice (taking $\tau=(2J_1,K-l,l)\  , l\in  Z\! \! \! Z_+ \ $)
in (\ref{G}) corresponds to lowest weight representations
at the first and the third points and highest weight representations
at  the second and the last. The correlator
is given by an expression in inverse powers of
$\ux$, times $\ux^{2J}$. Both the choice in \cite{D} and (\ref{s'=1})
lead to fusion transformations different from the
minimal ones. Namely they give a single contribution
$J_{j,j'+1/2}$ or $J_{j,j'-1/2}$ respectively, in the
product $J_{j,j'}\otimes J_{0,1/2}$ while both are present
in the minimal model fusion rules.
The absence of the second contribution is due to
non-meromorphic powers like the one in  (\ref{s'=1}) which ``wipes'' away
the same factor in the function $\,\Phi^{(0,1;\nu)}$.

This example shows that it is rather awkward to model
infinite dimensional representations of $g$ (which in general
are neither highest nor lowest weight representations)
by bosonic powers. So
instead of trying to give meaning to the bosonic expressions
we shall follow a different way. To understand our motivation
it is instructive first to consider the thermal case.

\section{Generalized hypergeometric integrals, the
        $(x-z)$--expansion and the reduction of the  KZ--ZF
         system of equations -- the thermal case }

\subsect{Equivalent integral representations of the thermal
         correlators}

Summarizing the formulae of the previous section, provided by
the bosonization construction, we have for the thermal $n$-point
correlators (\ref{G})

\BL{xb}
   G^{(n)}_{j}(x,z)=
    \  \sum_{|\tau|=s+2j_n} \  \prod_{a=1}^{n-1}
    {2j_a\choose\tau_a}  x_a^{2j_a-\tau_a}
    \sum_{|\mu|=s} \,\, \prod_{a=1}^{n-1} {\tau_a\choose\ma}\,
    I^{(s)}_{\mu \  \Gamma}(z)\  .
\EE

Since we will assume that we integrate over a fixed cycle
$\Gamma$ we will skip indicating it explicitly.
The integral $\ I_{\mu}\ $ is defined from the bosonization
scheme  for $\ |\mu|= s\ $. It can be extended to any
$\ \mu \,$, $\ 0 \leq |\mu| \leq s\,$, assuming that in (\ref{l})
$\ l_i=0,1 \ $  for any $i=1,2,...,s$ and summing over all  such
$\ \{l_i\}\ $, consistent with the condition $\ |l|=|\mu|\ $, i.e.,

\BL{averl}
  \ff_{\mu}(u,z)={(s-|\mu|)!\over s!}\sum_{l}\ff_{\mu}^l(u,z)\,.
\EE
If $|\mu|=s$ the r.h.s. of (\ref{averl}) contains one term
described by $l=\rho$ while for arbitrary $0\le |\mu|\le s$ it
reproduces (\ref{varphi}).

 We shall keep the same notation for the corresponding integrals.
 These generalized  hypergeometric integrals
satisfy the following relations  valid for  arbitrary parameters
$\ \nu \ne 0,\infty\ $
 and $\{ j_a, a=1,2,...,n\}$, such that $j_1+j_2+...+j_{n-1}-j_n=s$,
$s$ \? positive integer:

\vskip 1cm

the linear relation of Aomoto
\BL{tA}
  \sum_a(2j_a-\ma)I^{(s)}_{\mu+\ve_a}=0\, , \ \ (\ve_a)_b=\delta_{ab}\, ,
\EE

the recursion relation
\BL{tR}
  \sum_a(2j_a-\ma)\  z_a\  I^{(s)}_{\mu +\ve_a}
  =- c_{|\mu|} \ I^{(s)}_{\mu} \, ,
\EE

\BL{tc}
c_{|\mu|}=\sum_{a=1}^{n}j_a - |\mu| +1  -{1 \over \nu}\, ,
\EE

and the system of differential equations

\BAN
  \frac 1{\nu}\ \dz_a I^{(s)}_{\mu}&+&\sum_{b(\ne a)} \frac 1{z_{ab}}
  [\ma(2j_b-\mb)(I^{(s)}_{\mu} - I^{(s)}_{\mu-\ve_a+\ve_b}) +
  \mb(2j_a-\ma)(I^{(s)}_{\mu}-I^{(s)}_{\mu+\ve_a-\ve_b})]
\\ \label{tD}
  &=& (2j_a-\ma)(s-|\mu|)I^{(s)}_{\mu+\ve_a}\,,
  \qquad\quad a=1,\dots,n-1.
\EEA
In the case under consideration $\nu=1/(k+2)$ and
$0\le |\mu|\le s$. The relations (\ref{tA}) (for $|\mu|=s$) and
(\ref{tR}) were derived in \cite{A}. The equation  (\ref{tD}) is an
extension for $ |\mu|\le s$ of the corresponding equation
derived in \cite{A} for $ |\mu|= s$. The proof of these relations
will be reproduced in Sect. 4 as a particular case of a more
general computation.

Now we will show that the  conformal block (\ref{G2}) coming
from the bosonization can be rewritten in terms of a single
hypergeometric integral, namely
\BL{G3}
  {\cal G}_{(2j_1,\tau_2,\dots,\tau_{n-1})}(\uz) = (-1)^s
  I^{(s)}_{(0,2j_2-\tau_2,\dots,2j_{n-1}-\tau_{n-1})}(\uz)\,.
\EE
We have taken $\tau_1=2j_1$ since, using the isospin projective
invariance we can always choose $x_1=0$, so that in the
correlator (\ref{cor}) only the
component $m_1=-j_1$ , i.e., $\tau_1=2j_1$, survives.

To prove (\ref{G3}) we first iterate the Aomoto linear relation
(\ref{tA}) obtaining
$$
  I^{(s)}_{\mu} =
  { (-1)^{\mu_1 - \sigma_1} (\mu_1 - \sigma_1)! (2j_1 - \mu_1)!
   \over (2j_1 - \sigma_1)! }
  \sum_{\sigma_2,\dots,\sigma_{n-1}}
  \prod_{a=2}^{n-1}
  \left(\matrix{2j_a-\mu_a \cr 2j_a-\sigma_a }\right)
  I^{(s)}_{\sigma}
$$
where $|\mu|=|\sigma|=s$. Substituting this on the r.h.s. of
(\ref{G2}) with  $\tau_1=2j_1$, $\sigma_1=0$
 we obtain
$$
  (-1)^s \sum_{\sigma_2,\dots,\sigma_{n-1}}
  \sum_{\mu_2,\dots,\mu_{n-1}}
  \prod_{a=2}^{n-1}
  \left(\matrix{2j_a-\mu_a \cr 2j_a-\sigma_a }\right)
  \left( \matrix{\tau_a \cr \mu_a} \right) (-1)^{\mu_a}
  I^{(s)}_{(0,\sigma_2,\dots,\sigma_{n-1})}(\uz)
$$
This can be rewritten as
$$
  (-1)^s \sum_{\sigma_2,\dots,\sigma_{n-1}}
  \prod_{a=2}^{n-1} {(2j_a-\tau_a)! \over \sigma_a!}
  \prod_{a=2}^{n-1} {1 \over (2j_a-\tau_a-\sigma_a)!}
 I^{(s)}_{(0,\sigma_2,\dots,\sigma_{n-1})}(\uz)
$$
making use of the ${}_2F_1$ hypergeometry summation formula:
$$
  \sum_{\mu_a} (-1)^{\mu_a}
  \left(\matrix{2j_a-\mu_a \cr 2j_a-\sigma_a }\right)
  \left( \matrix{\tau_a \cr \mu_a} \right)
  = \left(\matrix{2j_a-\tau_a \cr \sigma_a }\right) \, .
$$
Note that if $A_a$ are integers and $\sum A_a=0$ then
$$
  \prod_a {1\over A_a!} = \prod_a \delta_{A_a,0} \,.
$$
Obviously
$$
  \sum_{a=1}^{n-1} (2j_a-\tau_a-\sigma_a) = 2(s+j_n)
  -|\tau|-|\sigma|=0
$$
so with the choice $\tau_1=2j_1$ and $\sigma_1=0$ we have
$  \sum_{a=2}^{n-1} (2j_a-\tau_a-\sigma_a) = 0 $
and thus
$$
  \prod_{a=2}^{n-1} {1 \over (2j_a-\tau_a-\sigma_a)!}
  =\prod_{a=2}^{n-1} \delta_{\sigma_a, j_a+m_a} \, .
$$
Therefore we obtain (\ref{G3}) which, inserted in (\ref{G}),
(\ref{G2}) gives

\BL{DJMM}
   G^{(n)}_{j}(\ux,\uz) = \  (-1)^s
   \sum_{\sg:\ {}^{|\sg|=s}_{\sg_1=0}}
   \  \prod_{a=2}^{n-1} {2j_a\choose\sg_a}
   \ \underline x_a^{\sg_a}\ I^{(s)}_{\sg}(\uz)\ .
\EE

The representation (\ref{DJMM}) corresponds to the one considered
in \cite{DJMM}. More precisely, inserting (\ref{DJMM}) in (\ref{cor})
and using the iterated Aomoto linear relation we can rewrite the
$x$ \? expansion for the correlator at infinity (\ref{W_inft}) as
\BL{U_inf}
  U^{(n)}(x,z)
  = \prod_{1\le a<b\le n-1} z_{ab}^{2j_aj_b\,\nu}\ G^{(n)}_j(x,z) =
  \sum_{|\al|=s} \prod_{a=1}^{n-1}
  {2j_a \choose \al_a} \, x^{\al_a}_a \,
  \prod_{1\le a<b\le n-1} z_{ab}^{2j_aj_b\,\nu} \ I_{\al}^{(s)} (z) \,.
\EE

Our next step will be to derive another equivalent
representation for the thermal correlators. As already argued in
the introduction  the 2- and 3-point functions of the WZNW model
reproduce in the limit $x_a \to z_a$ their minimal model
counterparts. We will show that this is true for arbitrary
$n$-point functions. Indeed we have the following $(x-z)$ \?
expanded form of the correlators

\BL{x-z}
   G^{(n)}_{j}(x,z)=  \ \sum_{t=0}^s
   \sum_{|\tau|=t} \prod_{a=1}^{n-1} (x_a-z_a)^{\tau_a}
   \, B_{\tau}\, I^{(s)}_{\tau}(z)
\EE
where

\BL{5.1.3}
   B_{\tau}=\frac  {(-1)^t }{(s-t)!} \prod_{r=0}^{s-t-1} c_{t+r} \
   \prod_{a=1}^{n-1} {2j_a\choose \tau_a}  , \qquad t=|\tau| \,.
\EE

The crucial element in the derivation of this formula is the
recursion relation (\ref{tR}). Iterating it $\,s-t\,$ times we obtain
\BL{5.1.4}
\sum_{|\sg|=s} \prod_{a=1}^{n-1} z_a^{\sg_a -\tau_a}
{2j_a-\tau_a \choose \sg_a -\tau_a} I^{(s)}_{\sg} =
(-1)^{s-t} \frac{\prod_{r=0}^{s-t-1} c_{t+r}}{(s-t)!} \
I^{(s)}_{\tau} \ .
\EE
Now expand
\BL{5.1.5}
  x_a^{\sg_a} = \sum_{\tau_a} {\sg_a\choose\tau_a}
  (x_a-z_a)^{\tau_a} z^{\sg_a-\tau_a},
\EE
and use the trivial identity for binomial coefficients
\BL{5.1.6}
  {2j_a\choose\sg_a}{\sg_a\choose\tau_a} = {2j_a\choose\tau_a}
  {2j_a-\tau_a\choose\sg_a-\tau_a}
\EE
to rewrite $\,G^{(n)}_j(x,z)\,$ in (\ref{U_inf}) as

\BL{5.1.7}
  \sum_{t=0}^s (-1)^t \sum_{|\tau|=t}
  \prod_{a=1}^{n-1} {2j_a\choose\tau_a} (x_a-z_a)^{\tau_a}  \left[
  \sum_{|\sg|=s} \prod_{a=1}^{n-1} z_a^{\sg_a -\tau_a}
  {2j_a-\tau_a \choose \sg_a -\tau_a} I^{(s)}_{\sg} \right] \ .
\EE
Thus, applying the iterated recursion relation for the
expression in the brackets gives us the formula for the $(x-z)$ \?
expansion.

The expansion (\ref{x-z})  simplifies when written
for the 4-point function in the system $(0,1,\infty )$
\BL{x-z4}
   G^{(4)}_{j}(\ux,\uz)=  \ \sum_{t}
   (\ux-\uz)^{t} \ C_t(\uz) =   \sum_{t}
   (\ux-\uz)^{t}\ B_{t}\ I^{(s)}_{t}(\uz)\, ,
\EE
where for short we have replaced the vector $\tau=(0,t,0)$ with
the number $|\tau|=t$.
The Aomoto differential equations (\ref{tD}) transform to
 a  system of ordinary differential
equations for the  integrals $I^{(s)}_{t}(\uz)$
\BL{ZF}
   ( (k+2) \dz_z+a_{tt}\, )\ I^{(s)}_t(\uz)
   = (2j_2 - t)\, (s-t)\ I^{(s)}_{t+1}(\uz)
   -{t \ c_t\over z(z-1)}\, I^{(s)}_{t-1}(\uz)      \ ,
\EE
\BL{a}
    a_{tt}= {A_t(j_1+j_2) \over z}+{A_t(j_3+j_2) \over z-1},
    \qquad A_s(j)=s(2j-s+1) \ ,
\EE
$$
  t=0,1,2,..., \min(s, 2j_2)\, .
$$

To show this we use the recursion relation (\ref{tR})
\BL{3.1.21}
  2j_3 I^{(s)}_{(0,t-1,1)} + z (2j_2 - t + 1) I^{(s)}_{(0,t,0)}
  +c_{t-1} I^{(s)}_{(0,t-1,0)} = 0
\EE
and the Aomoto linear relation (\ref{tA})
\BL{3.1.22}
  2j_3 I^{(s)}_{(0,t-1,1)} + (2j_2 - t + 1) I^{(s)}_{(0,t,0)} +
  2j_1 I^{(s)}_{(1,t-1,0)} = 0\ .
\EE
We solve for $I^{(s)}_{(1,t-1,0)}$ and
$I^{(s)}_{(0,t-1,1)}$ and
substitute in (\ref{tD}) which in this case (n=4) is simply
($\nu= 1/(k+2)$ )

\BA{tD4}
  (k+2)\dz_z I^{(s)}_{(0,t,0)} &=&
  (2j_2 - t)(s-t) I^{(s)}_{(0,t+1,0)}
\\ \nonumber
&&  - 2j_1\, t\, {1 \over z}\,(I^{(s)}_{(0,t,0)} -
          I^{(s)}_{(1,t-1,0)}) -
   2j_3\, t\, {1 \over (z - 1)}\,(I^{(s)}_{(0,t,0)} -
          I^{(s)}_{(0,t-1,1)})\, .
\EEA

 We get (\ref{ZF}) which is nothing but the  KZ equation
(\ref{KZ}) applied to the 4-point function (\ref{cor}) with
$\Omega_{ab}$ given in general by
\BL{Omega}
  \Omega_{ab} = - x_{ab}^{\,2}\, {\dz\over\dz x_a}\,
  {\dz\over\dz x_b}
  + 2 x_{ab}\left( J_a {\dz\over\dz x_b}
  - J_b  {\dz\over\dz x_a} \right) + 2J_a J_b \, .
\EE
Written in terms of the coefficients $\,C_t(z)=B_t\,I_t^{(s)}(z)\,$
in the $(x-z)$ \? expansion of the correlator (\ref{x-z4}),
the equation

\BL{dC}
   [(k+2)\partial_z + a_{tt}]C_t(\uz) + (t+1)\,c_t \ C_{ t+1}(\uz) +
   { (s-t+1)(2j_2-t+1) \over z(1-z)}\ C_{t-1}(\uz) =0\ ,
\EE
first appeared in \cite{ZF}. Here we have found
the solution in terms of the chiral integrals $I^{(s)}_{t}(z)$
and the explicit numerical coefficients $B_t$.
In \cite{ZF} a somewhat different parametrization of
the (two-dimensional) integrals is used with $\ \{j_a \}\ $
replaced by (see also \cite{ChF})

$$
\tilde {\jmath}_1 = {1 \over 2}(j_1+j_2-j_3-j_4+k+1) \,,
$$
$$
\tilde {\jmath}_2 = {1 \over 2}(j_1+j_2+j_3+j_4-k-1) \,,
$$
$$
\tilde {\jmath}_3 = {1 \over 2}(j_2+j_3-j_1-j_4+k+1) \,,
$$
$$
\tilde {\jmath}_4 = {1 \over 2}(j_2+j_1-j_3-j_4+k+1) \,.
$$
This transformation relates  the numerical coefficients in
(\ref{ZF}) and (\ref{dC}) so that up to an overall constant,
which can be computed using the results of \cite{DF}, we can make the
identification

\BL{I'}
   I_t^{(s)}(\{\tilde {\jmath}_a\};z)={\rm const}\ (s-t)!\,t!\,(-1)^t\,
   C_t^{(s)}({j_a};z) \ .
\EE

It is straightforward, using the relations (\ref{tA}), (\ref{tR}),
to show the equivalence, for arbitrary $\,n$,
 of (\ref{tD}) and the KZ \? ZF system of
equations written using (\ref{Omega}) for
$\,\{I^{(s)}_{\tau}(\uz),\, |\tau|\le s,\, \tau_1=\tau_{n-1}=0\}$.

Both the $x$ \?\, and $(x-z)$ \? expansions are finite series
in agreement with the algebraic equation
corresponding to a  singular vector of the $\ A_1^{(1)}$ \? Verma
modules. In the thermal case under consideration
it is simply a monomial and can be
 represented by the derivative of order
$2j_a+1$ with respect to $x_a$ -- exactly  as in the
integer level ($k+2=p; p'=1$) case. In the integer level
case  the second algebraic equation is reflected in  the
truncation from below of the $(x-z)$ \? expansion if
$c_0 \geq 0\,,$ integer \cite{ZF}, \cite{FSV}.
Indeed, examining the  numerical
coefficients in the expansion (\ref{x-z4}), the lower bound in
(\ref{x-z4}) is found to be $t_{\min}=\max (0,c_0+1)$ for $p'=1$.
For general $k$ one expects instead a linear relation for
the integrals in the $(x-z)$ \? expansion (see also Section 4.3).

The expansion (\ref{x-z4}), and more generally (\ref{x-z}), makes
 explicit the reduction of the WZNW correlators
to the minimal theory ones in the limit $x_a \rightarrow z_a$. What
we have done (for $n=4$)
 was to change the set of integrals entering the $x$ \?
 expansion (\ref{DJMM}),   provided by the bosonization,

\BL{Imu}
\{  I^{(s)}_{\mu},\  |\mu|=s\}\, ,
\EE
subject of the  linear relation (\ref{tA}), to
 an equivalent (for
generic isospin values) set

\BL{It}
\{  I^{(s)}_t, t =0,1,...,s\}\, ,
\EE
containing the minimal integral $I_0$
as one of its members.

This was done exploiting the recursion relation (\ref{tR})
which can be equivalently rewritten as
\BL{XR}
  \sum_{a=1}^{n-1} z_a\,\dz_{x_a}\, U_t(x,z) = U_{t-1}(x,z)
\EE
where $U_t(x,z)$, $t=0,1,\dots,s$ is defined by the r.h.s. of
(\ref{U_inf}) times an overall constant $(-1)^tc_t!/c_s!$ and
 the condition $|\al|=s$ replaced by $|\al|=t$. Applying (\ref{tR})
to the correlator $U^{(n)}(=U_s)$,
rewritten using (\ref{x-z}) as a $(x-z)$ \?
expansion in terms of $U_t(x-z;z)$, we obtain
\BL{XR2}
  \sum_{a=1}^{n-1} z_a\,\dz_{x_a}\,U^{(n)}(x,z) =
  \sum_{t=t_{\min}}^{s-1} {1\over(s-t-1)!} U_t(x-z;z)
\EE
while
\BL{XR3}
  s\,U^{(n)}(x,z) = \sum_{t=t_{\min}}^s
  {s\over(s-t)!}\,U_t(x-z;z)
\EE
For $t_{\min}=0$ the right hand sides of (\ref{XR2}) and (\ref{XR3})
coincide (yielding the Virasoro correlator $U_0$) if
$x_a=z_a$, $a=1,\dots,n-1$. Thus the condition
$U^{(n)}(z,z)=U_0(z)$ is equivalent
to the requirement that the operator in the l.h.s. of (\ref{XR2})
reduces to the identity operator. This relates the limit $x_a\to z_a$ to
the standard constraint condition of the quantum hamiltonian reduction
\cite{BO} since the operator $\Delta(X^-_1)\equiv \sum_a z_a\,\dz_{x_a}$
corresponds to the KM generator $X^+_{-1}$ applied to the vacuum state
created by the field $\Phi^{J_n}(x_n,z_n)$ at infinity.
In the integer level case
the vanishing of the $(k-2j_n+1)^{\rm th}$ power of $\Delta(X^-_1)$
applied on $U^{(n)}$ accounts for the algebraic equation mentioned above,
responsible for the truncation of the $(x-z)$ \? expansion from below.
Similarly the Aomoto linear relation (\ref{tA}) (for $|\mu|-1=s$),
reexpressed as a relation for $U^{(n)}$, $\sum \dz_{x_a}\,U^{(n)} =0$,
admits an algebraic interpretation  \cite{DJMM}.

The relations (\ref{tA}) and (\ref{tR}) can
be interpreted as coming from relations in twisted cohomology, that
is to say,   certain forms are exact  with respect to a twisted total
derivative $\nabla_{\Phi}=\nabla-\nabla ( log
\Phi )$ .  In \cite{A} it was shown that
$\varphi_{\mu}du_1\wedge...\wedge du_s$ with $|\mu|=s$ span the
symmetric top\footnote{ Symmetric with respect to the action of the
permutation group $\Sigma_s$ on the screening variables
$u_i$ $i=1,\dots,s$; top refers to $r$-forms with $r=s$}
twisted cohomology $[H^s(X_s,\nabla_{\Phi})]^{\Sigma}$
and when the Aomoto linear relations are taken into account one gets
a basis.
Its dimension is $\ (s+n-3)!/ (s!(n-3)!)\ $
for generic spins. The two sets (\ref{Imu}) and
(\ref{It}) (for $n=4$)
correspond to two different bases in cohomology.

Up to now we have considered a fixed cycle $\Gamma$. For non-rational
$k+2$  the general
solution of the KZ equation is given by a fundamental set of integrals
labelled by
 the cycles in the  dual twisted homology group. In the minimal
rational case, i.e., $2j\ $ restricted in the domain (\ref{jj'}), we shall
restrict ourselves to an ``admissible''  set of contours, which provides
a linearly independent set of integrals. Then
given the chiral solution (\ref{x-z}) one can construct  the
monodromy invariant 2 -dimensional correlations
 combining
the left and right chiral pieces. Let us consider in more detail
the case $n=4$. There are (for generic spins) $s+1$ cycles $\Gamma_p, \
p=1,2,..., s+1$, any $\,\Gamma_p\,$ describing a set of contours.
One can choose these sets of contours and the branches of the
function $\,\Phi^{(s)}$ as in \cite{DF}.
Then $\{I_{t;p}^{(s)}\}$ transform with the same fusion matrix as
$\{I_{0;p}^{(s)}\}$, times a sign $(-1)^t$. This can be seen using
the invariance of the system (\ref{ZF}). The sign is compensated
in the power series (\ref{x-z4}) when $x$ and $z$ simultaneously
go to $(1-x)$ and $(1-z)$, so that the full  function (\ref{x-z4})
transforms with the same minimal theory fusion matrix. Then the
construction of the monodromy invariant correlators repeats the
one for the minimal models.
Furthermore one can construct ``mixed'' monodromy invariants
combining the $(x,z)$ \? depending chiral correlators of the
WZNW theory with the corresponding $\overline z$ \? dependent
minimal conformal blocks.

\subsect{Reduction of the thermal KZ  system to the BPZ
            equations}

Let us start with the reduction of the KZ \? ZF system
for the case $n=4$. It is a first order matrix differential
equation for the coefficient functions of the $(x-z)$ \?
expansion of the correlators. This matrix differential
equation can be reduced to a scalar (BPZ) equation for the
correlators of the minimal models -- the latter reflecting
the decoupling of the Virasoro null vectors.

It is convenient to rewrite the  KZ \? ZF
system (\ref{ZF}) as follows
\BL{KZZF2}
   (2j_2-t)(s-t)I_{t+1}=D_{t+1} I_t +
   \frac{t\,c_{t-1}}{z(z-1)} I_{t-1}\ ,
\EE
where
\BL{Lt}
   D_{t+1} = {1\over \nu} \partial_z +
   t\left(\frac{2j_1+2j_2+1-t}{z}
   +\frac{2j_2+2j_3+1-t}{z-1}\right)
\EE
Using (\ref{KZZF2}) one may express recursively $I_{t}$
by the lower ones and finally represent it as a
differential operator of order $t$ acting on $I_0$:
\BL{BPZ}
   {(2j_2)!\over (2j_2-t)!}{s!\over (s-t)!}I_{t} =
   \sum_{k=0}^{[{t \over 2}]}
   \sum_{\{i_k,\dots,i_1\}}
   D_{t}\dots\overbrace{D_{i_k+1}D_{i_k}} D_{i_k-1}
   \dots\overbrace{D_{i_1+1}D_{i_1}} \dots D_2D_1 \,\, I_0
\EE
where the ``pairing '' is given by
\BL{}
  \overbrace{D_{r+1}D_r} \equiv (2j_2+1-r)(s+1-r)
   \frac{r\,c_{r-1}}{z(z-1)}\
\EE
and the second sum in (\ref{BPZ}) is over all subsets
$\{i_k,\dots,i_1\}$ of
$\{t-1,\dots,2,1\}$ with $i_{j+1}-i_j\ge 2$ and
 $[t/2]$ is the integer part of $t/2$. The $k=0$
contribution in (\ref{BPZ}) does not contain any
pairings.

In the thermal case, which we are considering now,
the KZ \? ZF matrix equation is finite and we can carry
out its reduction immediately.
Indeed, consider the relation (\ref{BPZ}) -- if we
set $t=s_0 \equiv \min (2j_2, s)$, then the left-hand side
vanishes and we get a scalar differential equation for the
integral $I_0$.

An explicit expression for the Virasoro singular vectors
in the  thermal case was written down in \cite{BsA}. Thus
via the conformal Ward identities one can immediately
write down the BPZ equations.
The ordinary differential equation obtained from (\ref{BPZ}) is the
null vector decoupling equation for the 4-point function when projective
invariance is taken into account, as can be checked immediately in the
simplest examples. In fact writing the BPZ
equation for the 4-point function as an ordinary differential
equation, though straightforward, becomes quite cumbersome as the
order of the equation grows. In this respect we may say that the
BPZ equation written in the form of (\ref{BPZ}) is more explicit.
An example of the reduction of the KZ equation has been also discussed in
the context of the coset construction in \cite{CKPS}.

The above consideration is not
restricted to 4-point functions. Dealing with an
arbitrary $n$-point function it is more convenient to send only the
last point to infinity instead of working with moduli.
We shall reduce the set of equations (\ref{tD}) which was shown to be
equivalent (using the linear and recursion relations (\ref{tA}),
(\ref{tR})) to the KZ \? ZF system.
Using (\ref{tD}) we can recursively
express $I_{\mu}$ as some differential operator acting on $I_0$.
Doing this in the ``$a^{\rm th}$-direction'', $\,a=1,\dots,n-1\,$, i.e.,
expressing  recursively $I_{t\,\ea}^{(s)}$, $t=0,1,\dots\,$
we obtain that $I_{(s_a+1)\varepsilon_a}^{(s)}$ multiplied by
a vanishing coefficient is given by a differential operator
 acting on $I_0^{(s)}$. (Here $\,s_a=\min (2j_a,s)\,$ and we shall
assume for simplicity that $\,2j_a=s_a\,$ for some $\,a\,$.)
Then the analogue of (\ref{BPZ}) in the case of general $n$ (with only
$z_n$ fixed at infinity) is
\BL{bpzn}
    {I\!\!i}_t = {\cal N}_t(j_a)\  {I\!\!I}_0 -
   \sum_{l=1}^t \frac{\nu^l}{l!}
   \left( \prod_{i=0}^{l-1}(t-i)\,(2j_a+1-t+i)\right)\,
   \left(\left( \sum_{{}^{b=1}_{(b\ne a)}}^{n-1}
   \frac{j_b}{z_{ab}}-\da\right)^l\cdot 1\right)
   \ {I\!\!I}_{t-l}  \,.
\EE
We have set
$$
   {I\!\!I}_{\mu} = \nu^{|\mu|}\,\frac{s!}{(s-t)!}\,
   \frac{\prod_a (2j_a)!}{\prod_a(2j_a-\ma)!}
   \prod_{1\le a<b\le n-1} z_{ab}^{2j_aj_b\nu} \ I_{\mu}
$$
and for short $ {I\!\!I}_t= {I\!\!I}_{t\ea}$ while
${\cal N}_t(j_a)$ is a differential operator of order $\,t$, e.g.,
${\cal N}_1=\da$,

\BL{}
   {\cal N}_2(j_a)=\da^2-\nu\,2j_a\,{\cal L}_{-2}\,,
\EE
\BL{}
   {\cal N}_3(j_a)=\da^3-\nu\,(2j_a\,\da\,{\cal L}_{-2} +
   (2j_a-1)\,2\,{\cal L}_{-2}\,\da)
   +\nu^2\,2j_a\,(2j_a-1)2\,{\cal L}_{-3}\,, \quad {\rm etc.,}
\EE
 with
$$
  {\cal L}_{-m} = (-1)^m\sum_{b\neq a}^{n-1}
\frac{1}{z_{ab}^{m-1}}
\left( \frac{(m-1)h_b}{z_{ab}}+\partial_b \right)
$$
being the standard BPZ differential operator
realization for the Virasoro generators \cite{BPZ}.

For $\,t=2j_a+1\,$ the l.h.s. of (\ref{bpzn}) and the second term
on the r.h.s. vanish, while ${\cal N}_t(j_a)$
becomes the operator representing the respective Virasoro
 singular vector.  Hence one recovers the BPZ equation
arising from the decoupling of this  vector -- given by a
descendant of the primary field sitting at the point $\,z_a$.

The form (\ref{bpzn}) in which we have cast the KZ system of
equations makes explicit its connection with the system written
in \cite{BFIZ}, namely, for any $t$ the first term on the
r.h.s. of  (\ref{bpzn}) can be identified with a state
in the $sl(2)$ module built there. Thus the two systems
are related by a
kind of Drinfeld \? Sokolov ``gauge transformation''.
This relation and the derivation of (\ref{bpzn})
will be presented in more detail elsewhere.

\sect{Rational level and isospins $A_1^{(1)}$ model. }
\subsect{Generalized hypergeometric integrals -- the
``meromorphic'' solution.}

We have written basically two different integral  representations
for  the thermal correlations, the $x$ \?  and $(x-z)$ \? expansions,
and we have shown that they are equivalent.
 The attempts to extend the first
expansion, intrinsically related to the bosonization scheme,
for general rational isospin values
lead, as we have seen in Sect. 2.2, to non-meromorphic
modification factor for the integrand of the  minimal theory.
On the other hand we can try to find directly a solution of the
general KZ equation, which  is obtained by inserting in (\ref{KZ}) the
 isospin $\ sl_I(2,C\!\!\!\!I)\ $ generators. Assuming that the
solution is represented as a power series in $(x-z)$ we get an
infinite system which
 looks like the ZF system for the coefficients $C_t$ (with
 $j_a$ replaced by $J_a$),  only
now it is not restricted from above. We will now describe
solutions
with coefficients characterised by a meromorphic modification
function.

Let $\mu, \tau$ and $\sg$ be $(n-1)$-vectors with
non-negative integer components. Set
\BL{Lphi}
   \varphi_{\mu}^{[L]}(\{u_i,z_a\}) = {(
   L!)^{m} (mL- |\mu|)! \over
   (mL)!} \sum_{\{N_{ia}\}}{\prod_a \mu_a! \
   \over \prod_i[( L-\sum_a
   N_{ia})! \prod_a  N_{ia}!(u_i-z_a)^{ N_{ia}}]} \ .
\EE
Here  the sums are finite
running over $\ N_{ia}=0,1,...,\mu_a\ ,$ $i=1,2,...,m\ , \
a=1,2,...,n-1, \ $  subject
to the constraints
\BL{Na=ma}
   \sum_{i=1}^{m} N_{ia} = \mu_a \,.
\EE
It is not difficult to show that for L=1 and
$\ 0 \le |\mu|\le s$
(\ref{Lphi}) reduces to the thermal factor (\ref{ff}).
Next define
\BL{nphi}
   \varphi^{(s,s')}_{\sg}(\{u_i,v_{i'},z_a\})={ (s'K)!s! \over
   S!}\sum_{\tau
   }\prod_a{\sg_a \choose \tau_a} {S-|\sg| \choose
   s-|\tau|}
  \varphi_{\sg-\tau}^{[K]}(\{v_{i'},z_a\})
  \varphi_{\tau}^{[1]}(\{u_i,z_a\}) \ .
\EE
Here $K\equiv -k-2$, $S=s+s'K$ and the sum  runs
over $\{\tau_a=0,1,...,s,\quad |\tau| \leq s\}$.
The expression (\ref{nphi}) is the general non-thermal factor.
It reduces to the thermal one when $s'=0$ or the quasithermal
$\ff^{[K]}$
when $s=0$. The convention will be that $m$ from (\ref{Lphi})
is $s$ in the thermal case $L=1$ while in the quasithermal case
$L=K$ it should be identified with $s'$ and the screening
variables are $v_{i'}$ instead of $u_i$.

The object of our interest are the generalised hypergeometric
integrals
\BL{nint}
   I_{\Gamma,\sg}^{(s,s')}=\int_{\Gamma} du_1\dots du_s
   \,  dv_1\dots
   dv_{s'} \, \,  \ff_{\sg}^{(s,s')} \, \Phi_{{J}}^{(s,s';\nu)}\, .
\EE
where we have used the general multivalued integrand
(\ref{PhiPhi}) of the  minimal correlations
and $J$ is an $(n-1)$ \? vector  $\ J=(J_a=j_a+j'_a\,K)\ $, while $\ s, s'\
$ are given in (\ref{ss'}).
It is assumed that the set of contours $\Gamma$ represents a
cycle in the relevant twisted homology group
so that no boundary terms survive.
Our main result is summarized in the following

\vskip 1cm \noindent
{\bf Proposition.}\, {\it The integrals
(\ref{nint}) satisfy  the following relations:}
\vskip .5cm

{\it the linear relations  }
\BL{nA}
  \sum_a(2J_a-\sg_a)I^{(s,s')}_{\sg+\ve_a}=0,
\EE

{\it the recursion relations}
\BL{nR}
  \sum_a(2J_a-\sg_a)\  z_a\  I^{(s,s')}_{\sg +\ve_a}
  +c_{|\sg|} \ I^{(s,s')}_{\sg} = 0\, ,
\EE

\BL{nc}
c_{|\sg|}=\sum_{a=1}^{n}J_a - |\sg| -k-1\, ,
\EE

{\it and the differential equations}

\BAN
  (k+2)\ \dz_a I^{(s,s')}_{\sg} &+&
   \sum_{b(\ne a)} \frac 1{z_{ab}}
  [\sg_a(2J_b-\sg_b)(I^{(s,s')}_{\sg} - I^{(s,s')}_{\sg-\ve_a+\ve_b})
  + \sg_b(2J_a-\sg_a)(I^{(s,s')}_{\sg}-I^{(s,s')}_{\sg+\ve_a-\ve_b})]
\\ \label{nD}
 &=& (S-|\sg|)(2J_a-\sg_a)I^{(s,s')}_{\sg+\ve_a} \,.
\EEA

\vskip 1cm
This Proposition holds true for arbitrary $k+2\neq 0$ and
any combination of $\{ J_a=j_a-j_a'(k+2)\} $ producing,
according to (\ref{ss'}), non-negative integers $s, s'$.
Its  proof  is rather technical and the whole
of Section 5 is devoted to it.

Repeating the argumentation of the previous section one can show
that the system of equations (\ref{nD}) can be rewritten as the (now
infinite)  KZ \? ZF type system.
Thus we immediately obtain that
the correlation functions $W^{(n)}_{J}(\{x_a,z_a,J_a\}_{a=1}^n)$
(cf. (\ref{cor})) with
\BL{x-zn}
   G^{(n)}_{J}(x,z)= \sum_{t=0}
   \sum_{\tau:\,\,
   |\tau|=t,}  \prod_{a=1}^{n-1}
   ( x_a- z_a)^{\tau_a}
   B_{\tau}\ I^{(s,s')}_{\tau}( z)\ ,
\EE
where the integrals $I^{(s,s')}_{\tau}( z)$ are defined
in (\ref{nint}) with the ``meromorphic'' $\ff^{(s,s')}_{\tau}$
 from (\ref{nphi}) and
\BL{B_tau2}
   B_{\tau}=\frac{(-1)^t\,\beta_{J}}{(S-t)!}\frac{(c_t)!}{(c_S)!}
  \ \prod_{a=1}^{n-1} {2J_a\choose \tau_a},\quad t=|\tau|\, ,
\EE
are solutions of the KZ equation.
They have the same braiding properties as their Virasoro
model counterparts which are obtained when we set $x_a=z_a$.
More precisely the limit $\,x_a\to z_a\,$ exists if
$\,B_0\ne 0$. In (\ref{B_tau2}) $\,\beta_{J}\,$ is an overall
constant chosen in such a way that all $\,B_{\tau}\,$ are
finite. It is necessary because the relative constants
$\,B_{\tau}/B_0\,$ may blow up.

In particular, for values of $k$ and $J_a$ as in (\ref{kJ}), (\ref{rat}),
(\ref{jj'}) and choosing an admissible (``physical'') set of
cycles,  we get a set of functions
which transform in a way consistent with the general minimal
models fusion rules, i.e., the multiplicities $N^{J(j,j')}
_{J(k_2,k'_2)\, J(k_1,k'_1)}$ are given by the same factorized
expression as the r.h.s. of (\ref{N=NN}).

The integrals (\ref{nint}) can be found directly -- at least for
the first several values of $|\mu|$ -- requiring that the
relations (\ref{nA}), (\ref{nR}) hold. This fixes the factor
(\ref{Lphi}) uniquely providing two solutions -- one for $L=1$
recovering the thermal integrals and another for $L=K$.
Note that the relations (\ref{nA}), (\ref{nR}) and hence the
functions (\ref{Lphi}), corresponding to these two ``one-type
screening charges'' integrals are different (one has a factor
$(2j'_a K-\ma)$ instead of $(2j_a-\ma)$). Then one checks that
the solutions satisfy the differential equation (\ref{nD}).
However it is technically rather difficult to proceed in this way
for higher values of $|\mu|$. Instead of that one can try to
follow and generalize the argumentation in the thermal case, where
the Aomoto relations (\ref{tA}),(\ref{tR}), provided the bridge
to the $\ (x-z)$ \? expansion. The idea is first to extend for
$\,|\mu|\le S=s+s'K\,$  the general bosonic expressions
(\ref{ff}), (\ref{phin})
  (assuming that $K\ $ and $\ 2J_a+1\ $ are positive integers)
and then do the analytic continuation to the rational values of these
parameters at a later stage.

Let $\z (r,m)$ consist of sets $\z =\{i_1,\dots,i_r\}$ with each
$i_k$ running over $1,\dots,m$ independently of the rest. Define
the $m$-vector $\zz=\sum_{i\in \z}\ve_{i}$,
$\,(\ve_{i})_j=\dl_{ij}\,$ thus
$\,(\zz)_j=\dl_{i_1j} +\dots + \dl_{i_rj}$ and $|\zz|=r$.
Introduce the function
\BL{phiL}
   \fLm = \frac{(L!)^m}{(mL)!}\sum_{\z\in \z (r,m)}\fIm,
   \qquad |\mu|=mL-r ,
\EE
where $\rho$ is the $m$-vector $(1,1,\dots,1)$ and $\,\ff^l_{\mu}\,$
is defined in (\ref{l}),  (\ref{sA=ma}),  (\ref{ff}).

For an $m$-vector $l$ with integer entries and such that $|l|=r$
the number of $\z\in\z(r,m)$ such that $\zz=l$ is equal to
$|l|! (\prod_i l_i!)^{-1}$. Hence we can rewrite
the above formula as

\BL{mphi}
\frac{(L!)^m}{(mL)!}   \sum_l {(mL-|\mu|)! \over \prod_i
l_i!}\,\ff_{\mu}^{L\rho-l} = \frac{(L!)^m}{(mL)!}
   \sum_{A}
   \frac{(mL-|\mu|)!}{\prod_i (L-\sum_a \iaa)!}\,
   \frac{\prod_a \ma!
\delta_{\ma , \sum_i A_{ia}} }{\prod_{i,a} \iaa! (u_i-z_a)^{\iaa}}.
\EE
Keeping $\ \{ \ma \} \ $ as  non-negative integers,  we
can analytically continue the last formula
to non-integer $L$,  recovering  (\ref{Lphi}).
Indeed, unlike $\ \{l_i\}\ $ in the constraint (\ref{l}), the
parameter $L$ enters (\ref{mphi})   only in the numerical
factors, not restricting the summation variables $\ A_{ia}\ $
(denoted by  $\ N_{ia}\ $in (\ref{Lphi}) ).  Note that the
meromorphic  solution (\ref{mphi}) can itself be represented as
an $n$-point bosonic correlation with one of the coordinates
taken at infinity (see \cite{FGPP}).

The factor (\ref{Lphi}) can be rewritten as

$$
  \varphi_{\mu}^{[L]}(u,z) = {(mL-t)! \over (mL)!}
  \sum_{ i_1,i_2, ..., i_t}C_{i_1 i_2...i_t}{1 \over
  u_{i_1}-z_{a_1}}...{1 \over u_{i_t}-z_{a_t}}
$$
where $\mu=\epsilon_{a_1}+...+\epsilon_{a_t},\ \ t=|\mu|$. The
constant $\ C_{i_1i_2...i_s}\ $ is symmetric with
respect to all its variables and has a non-vanishing value
for any partition  $P$ of $t$; e.g., two coinciding indices
correspond to a partition $\ (t=2+1+1+...+1)\ $, etc. In the
thermal case $\ L=1\ $ only the  terms with
non-coinciding $\ \{i_1,.., i_t\} \ $ survive, corresponding to
the partition $P=(t=1+1+...+1)$ of $t$.

We can think of our meromorphic solution  as generated from the
minimal integral $I_0^{(s,s')}$ via the KZ \? ZF equation.
In the non-thermal (or quasithermal) case this is
an   infinite matrix system thus there is a functional
arbitrariness in the solutions. Choosing the minimal model
correlators as ``initial conditions'' for the ``evolution in
the discrete time $t$'' given by the KZ \? ZF system in the form
(\ref{dC}) reproduces our  solution, i.e., the infinite
set of integrals $\ \{    I_{\Gamma,\sg}^{(s,s')} \} \  $,
serving as coefficients in the expansion  (\ref{x-zn}).

Note that there are other solutions of the general KZ equation
which do not have this property. E.g., a solution
exists representing $\,G^{(4)}(\ux,\uz)\,$ as a
power series in the inverse powers of $\,(\ux-\uz)\,$
times $\,(\ux-\uz)^{2J_2}$.
The integrals serving as coefficients in this series
are defined using non-meromorphic factors. For example,
in the simplest case $\,s'=1\,$, $\,s=0\,$ these are obtained by
formally replacing $\,t\,$ with $\,2J_2-t$.
We should also mention that the proof of the relations (\ref{nA}),
(\ref{nR}) and the equation (\ref{nD}) formally extends for integrals
defined using analytic continuation of the factors (\ref{phiL}) to
non-integer $L$ and ${\mu_a}$, with $r$ assumed to be a non-negative
integer. Then, as in the thermal case, the ``non-meromorphic''
analogue of the system (\ref{nD}) is truncated from above
for $|\sigma|=S$, or $\sigma_a=2J_a$.

\subsect{Duality transformation and the truncation of the
quasithermal KZ \? ZF system}

In the quasithermal case (i.e., $s=0,\ S=Ks'$)
the KZ \? ZF system is an infinite system.  For $\,n=4\,$,
as in the thermal case,
we can write down, replacing $\,j_a\,$ with $\,J_a\,$, the
analogue of (\ref{BPZ}) but this time the left-hand
side never vanishes and contrary to the thermal case we
do not get an equation of BPZ type for $I_0$
from the KZ \? ZF system alone. Instead
(\ref{BPZ}) only describes the higher $I_n$ in terms
of $I_0$.
On the other hand, on the level of the minimal theory there is
apparently no asymmetry between the ``thermal'' and
``quasithermal'' cases. The  transformation
\BL{dKJ}
  K \rightarrow  1/ K\, \qquad
  J\rightarrow J/  K =j'_a+j_a/  K\ ,
  \quad
\EE
which keeps invariant (\ref{h(J)-J}),
amounts to a transposition of the Kac table, i.e.,
exchanging the rows and columns of the Kac table, or
equivalently the ``+'' and ``$-$'' screening charges.
Thus we can start with another WZNW theory with ``dual''
level  and isospin values,
whose  correlations reduce to the same minimal
correlations.
In the quasithermal case under consideration the dual
set is finite.
Comparing the two KZ \? ZF systems of equations
we can express any of the integrals $\{I_t^{(0,s')}\}$ in the
infinite set in terms of this finite basis. This implies additional
relations for the quasithermal integrals, namely any of them
 for $t>s'$
will be expressed  as a  linear combination (with coefficients that in
general are rational functions of $z$)  of the subset
$\ \{I_t^{(0,s')}\}\ $ with
$\ t=0,1,\dots,s'\ $. In this way the equation
(\ref{BPZ}) effectively truncates and can be furthermore
reduced, producing again a
BPZ type equation for the DF integrals.
We shall present an explicit algorithm to obtain
the additional  relations for the quasithermal integrals.

If  $\FFF$ is  a matrix, denote by $\FFF^{(k)}$ the matrix
consisting of the $k$-th diagonal, i.e.,
$\FFF^{(k)}_{ij}=\dl_{j,i+k}\FFF_{i,i+k}$.
For short denote $\FFF^{(k)}_i = \FFF^{(k)}_{i,i+k}$.
The  KZ \? ZF system of coupled first-order in $z$ differential
equations involves at most three $t$'s , i.e., $I_t$,
$I_{t\pm1}$. It  can be written as the following first-order
matrix differential equation for the column vector
${\bf\rm I}^{(s,s')}=(I^{(s,s')}_t)$:
\BL{KZ-ZF}
   (\dz \, 1\!\!1 + \FFF)\ {\bf\rm I}^{(s,s')} = 0 \ ,
\EE
where $1\!\!1$ is the unit matrix,
$\dz=\dz/\dz z$ and $\FFF\,(=\FFF(J,K,S),\ S=s+s'K)$
is a tridiagonal matrix with

\BA{FFF1}
   \FFF^{(1)}_t(z) &= \FFF_{tt+1} &=
   \frac 1{K}(2J_2-t)(S-t)
\\ \label{FFF2}
   \FFF^{(0)}_t(z) &= \FFF_{tt} &= \frac {-t}{K}
   \left[\frac{2J_1+2J_2+1-t}{z}
   +\frac{2J_2+2J_3+1-t}{z-1}\right]
\\ \label{FFF3}
   \FFF^{(-1)}_t(z) &= \FFF_{tt-1} &=
   \frac{-t}{K}\frac {c_{t-1}}{z(z-1)}
\EEA
and all other diagonals zero. We recall
\BL{5c}
   c_t= \sum_{a=0}^4 J_a-t +1+ K.
\EE

For short let us denote by $I_t=I^{(s,0)}_t(J,K)$ and
$\FFF=\FFF(J,K,S)$ $\,(S=s)\,$ the thermal ones while
their quasithermal dual counterparts we will denote
by ${\hat I}_t=I^{(0,s)}_t(\hat{J},\hat K)$ and
$\hat\FFF=\FFF(\hat J, \hat K, \hat S)$ where $\hat K=1/K$,
$\hat J=J/K$ and $\hat S= S/K = \hat Ks$.
  Obviously
\BL{I0=I0}
   \hat I_0=I_0 .
\EE
We want to determine a lower triangular matrix
$\BBB$, i.e.
$\BBB=\BBB^{(0)}+\BBB^{(-1)}+\BBB^{(-2)}+\dots$,
(a  Drinfeld-Sokolov \cite{DS} ``gauge transformation'')
transforming $(I_t)$ into $(\hat I_t)$:
\BE
   \hat I_{t} = \BBB^{(0)}_t I_t + \BBB^{(-1)}_t I_{t-1}
   +\BBB^{(-2)}_t I_{t-2}+\dots+\BBB^{(-t)}_t I_0
\EE
and such that
\BL{BF=FB}
   \BBB(\dz + \FFF) = (\dz + \hat \FFF) \BBB ,
\EE
i.e., the effectively truncated equation is equivalent to
its dual. An algorithm for the recursive determination of
the entries of the matrix $\BBB$ is given in the Appendix.

Let us illustrate in the example
$s=1$ the linear dependence we were talking about, i.e., of the
quasithermal integrals $\hat I_t = I^{(0,1)}_t(\hat J,\hat K)$
only $\hat I_0$ and $\hat I_1$ are independent. We have that
$s=1$ implies
 $\BBB^{(0)}_2=0$ (see (\ref{5b1})),
 reflecting the finiteness of the
thermal system. Thus
$\hat I_2=\BBB^{(-1)}_2 I_1 +\BBB^{(-2)}_2 I_0$. We have
$\hat I_0 = I_0$. Also $\hat I_1 = I_1$ because $\BBB^{(0)}_1=1$
and $\BBB^{(-1)}_1=0$ (see above). Therefore
\BL{I210}
   \hat I_2=\BBB^{(-1)}_2 \hat I_1 +
   \BBB^{(-2)}_2 \hat I_0
\EE
with the coefficients given in (\ref{5b2}) and
(\ref{5b3}) with $s=1$.

This furthermore extends to $C_t(z)=B_t\, I_t^{(0,t)}(z)$ for
arbitrary $t$, i.e.,

\BL{conv}
C_{t+2}(z)
=a_{t+1}\,({1\over z}+{1\over z-1})\,C_{t+1}(z) +b_t\,{1
\over z(z-1)}\,C_t(z) \ \
\EE

$$
  a_{t+1}={(2K-t)\over(t+2)(3K-t)}\,, \qquad
  b_t={(K-t)^2(K-1-t)\over (t-3K) (t+1)(t+2)}\, ,
$$
where for simplicity we have chosen all $2J_a=K$, $a=1,\dots,4$.
This simple relation can also be derived directly using the
explicit expressions for the integrals $I_t^{(0,1)}$ and it is
equivalent to a standard relation for the solutions of the
hypergeometric equation.

The numerical coefficients in the recurrence linear relation
 (\ref{conv})  are finite in the limit $t \rightarrow \infty$.
This implies  that for any fixed $z, \  z \not = 0,1$ there
is a finite
region for $|x-z|$, namely

$${ |x-z|\over |z|}<1 ; \ \ { |x-z|\over |1-z|}<1 \, ,$$
in which the $(x-z)$ \? power series, representing the correlation
in this simple example, is
absolutely convergent.

We stress that the duality does not mean that there is a direct
relation between the correlators of the two theories. One is
described by finite and the other by
infinite-dimensional representations of the algebra
$sl(2,C\!\!\!\!I)$.
The quasithermal correlators can be
written as a finite $(x-z)$ \? expansion, but with coefficients
depending on $x$ and $z$.

The analysis above can be extended to arbitrary $n$-point functions
in a way analogous to what was done at the end of Section 3.2.
In particular we can compare the thermal system written in the form
(\ref{bpzn}) with its dual quasithermal analogue. Then the vanishing
of ${\cal N}_{2j_a+1}(j_a)$ (i.e., the thermal BPZ equation) is
equivalent to a linear relation involving the quasithermal integrals
$I_{\mu}$ with $|\mu|\le\min(2j_a+1,s)$.

\subsect{Truncation in the general non-thermal case}

The duality of the thermal and the quasithermal theories
established in the previous subsection has a natural
cohomological interpretation. The generalized hypergeometric
integrals which appear as coefficients in the $(x-z)$ \? expansion
are integrals of top twisted differential forms. Since the
cohomology is finite-dimensional, only a finite number of these
integrals are linearly independent (over the rational functions
of $z$). Indeed since the ``twisting''  function $\Phi_{\hat
J}^{(0,s;\hat \nu)} = \Phi_{J}^{(s,0; \nu)}\ $
($\hat \nu= 1/\nu = -K$) is identical in
both theories and  the factor $\varphi_t^{[K]}\ $ (\ref{Lphi}) for
any $t$ is a rational function, the  corresponding  form
$\ \varphi_t^{[K]}du_1\wedge \dots \wedge du_s$ can be
expressed (up to an exact form $\ \nabla_{\Phi} f\ $)  in terms
of a standard basis in the symmetric cohomology
$[H^{s}(X_s,\nabla_{\Phi})]^{\Sigma}$,
$\ \Phi=\Phi_{J}^{(s,0;\nu)}$.  In the case under consideration such
a standard basis is provided for generic isospins  by the
thermal (symmetrized) set (\ref{varphi}),
with $\ |\mu |=s,\ \mu_1=0 $.

Let us now turn to the general non-thermal case.
The situation here is more subtle. Indeed both the initial and
the dual theories are described by infinite-dimensional sets of
integrals and we cannot truncate the KZ \? ZF system exploiting this
duality. On the other hand, since our solution is again described
by a meromorphic function we can apply the cohomological
argumentation to show that there is a finite subset of basic
integrals, so that again the  system can in principle be
truncated. The general twisted cohomology
group  $\ H^{m}(X_m,\nabla_{\Phi}) $ \cite{A}, associated
to a   multivalued function of the type in (\ref{intPhi}) with
general (nonsymmetric) values of the exponents
$\,\{\lambda_{ia}, \lambda_{ij} \, ,\ i,j=1\dots,m;\
a=1,2,...,n-1\} $,  is spanned by a finite set of logarithmic
differential $m$-forms. The reader can find in \cite{A} a linear
relation for these $m$-forms generalizing the symmetric
(thermal) linear relation (\ref{tA}) as well as a generalization
of the differential system of equations (\ref{tD})
(a Gauss - Manin system) for the corresponding integrals.
The rank of this cohomology computed for generic values
of the exponents is $\ (m+n-3)!/(n-3)!\ $.

The non-thermal integrals with $m=s+s'$ and a set of
exponents given in (\ref{Phi}) can in principle be described
by this general twisted cohomology. Each of the factors in
(\ref{nphi}) corresponds to a rational form and hence can be
expressed in terms of the finite basis in \cite{A}. That would
imply the truncation of the KZ \? ZF equations.
On the other hand the multivalued function $\Phi_J^{(s,s';\nu)}$ in
the non-thermal integrals actually provides a very specific
example of the general ``twisting'' factor. Indeed, it is given
by an expression that is a product of two thermal (depending on
$u_i$ and  $v_{i'}$ respectively) factors that
are ``coupled'' by integer exponents $\lambda_{i,i'}=-2$.
That can lead to the effective reduction of the number of
the basic integrals (at least on an
 ``admissible'' subset of contours) to the  product of the
ranks of the corresponding symmetric cohomologies.
(Here, for simplicity, we assume
 that the spins  $\{j_a, j'_a\}$ are small enough in comparison
with $ \ p,p'$.)

Let us illustrate this by a simple example for $\,n=4\,$
and $\,m=s+s'=2$. We can
integrate  any $\Phi^{(1,1;\nu)}$ \? twisted
2-form $\,\varphi(u,v)\,du\wedge dv$ with
respect to one of the variables, say $\,u\,$, over a closed contour
around the second variable $\,v$. This defines an operator ${\cal
J}$ from  the cohomology $H^2(X_2,\nabla_{\Phi})$ (of rank six)
to the cohomology $H^1(X_1,\nabla_{\tilde\Phi})\,$ (of rank two),
where $\Phi=\Phi_J^{(1,1)}(u,v))$,
$\, \tilde\Phi=\Phi_{\tilde{J}}^{(1)}(u))$,
$\, \tilde J=J(K+1)/K$. The range of ${\cal J}$ is 2 \?
dimensional hence its kernel  is 4 \? dimensional.
 For example, if $\varphi_{\mu}
=\varphi_0 $, the integral over $u$ above reduces to the unique
linear relation (\ref{tA}) for the 2 \? dimensional cohomology.
 One
can check inductively, starting from $\varphi_0$ (i.e., from
the minimal integral) and using the KZ equation, that any of the
meromorphic forms in the infinite set of integrals
$\{I_{(0,t,0)},\  t=0,1,2\dots\}$ is in the kernel of this
operator.

Although this interpretation of the non-thermal theory explains
 in principle the truncation of the KZ-ZF system, it is not easy to
perform it explicitly. Indeed (following the strategy in the quasithermal
case), in order to relate the infinite KZ \? ZF system and the finite system
in \cite{A} -- written for different sets of integrals -- we first have to
rewrite the latter in another finite basis, which includes the minimal
integrals $I_0^{(s,s')}$.

 On the other hand there is  another
more direct mechanism for  finding explicitly the relevant
additional relations  for the coefficients in the
$(x-z)$ \? expansion. Namely we can  take into account
 the algebraic equations for the primary field correlation
functions, corresponding to the singular vectors of the KM Verma
modules \cite{KK}.

We recall that in the integrable case $\ k+2=p\ $ there are two
generating equations, which are differential equations
in the functional $x$-realization. They correspond to monomials
of the $\ A_1^{(1)}\ $ generators $f_1=X_0^-$ and $f_0=X_{-1}^+$
acting on the vacuum,  of degrees $ 2j+1$ and $p-2j-1$
respectively.  In the case of rational $\ k\ $ the structure of
the  reducible  $\ A_1^{(1)}\ $ Verma modules $\ V(k,J_{j,j'})\
$ is  essentially of the same type. However
the corresponding singular vectors labelled by  $J_{-j-1, j'}$
and $J_{p-j-1,j'}$ ( the first two in an infinite sequence) are
realized in general by homogeneous  polynomials
$\ {\cal P}_{t_1,t_0}(f_1,f_0)\,v_0\ $ of both
$f_1$ and $f_0$, of  degrees

\BL{s1}
 t_1=(2j+1)(2j'+1),  \qquad  t_0= (2j+1)2j'\ ,
\EE
and

\BL{s2}
t_1=(p-2j-1)(p'-2j'-1), \  t_0= (p-2j-1)(p'-2j')\ ,
\EE
respectively. The first of these vectors is also present
for arbitrary $k+2\neq 0$ and isospin $J_{j,j'}$ as in
(\ref{kJ}).

The general formulae for these polynomials \cite{MFF} look rather
unexplicit. For an illustration we shall use a simple explicit
formula presented in \cite{Dob}
which holds for the cases $t_1=2t_0$
or $t_0=2t_1$ respectively.
Namely the singular vector corresponding
to (\ref{s1}), i.e., when $2j'+1=2$, $j$ is arbitrary, reads

\BL{VV}
   v_s=\sum_{s=0}^{2m}(-1)^s{2m \choose s}{m_1 \over
        m_1-s}(f_1)^{2m-s}(f_0 )^m (f_1)^s v_0
\EE
where $m_1=2J+1,\  m=m_1-K \ (=2j+1)$.

 For $t_0=2t_1$ the same formula can
be used for the case in (\ref{s2}) (i.e., $2j'+1 = p'-1$,
$\,m=p-2j-1$ -- arbitrary), interchanging in
(\ref{VV}) $\,(m_1, f_1)\,$ and
$\,(m_0,f_0)\,$, where $\,m_0=k+2-m_1$. Our convention for the
$A_1^{(1)}$  commutation relations is
\BL{KMcr}
  [X^{\al}_n,X^{\bt}_m] = f^{\al\bt}_{\ga} X^{\ga}_{n+m}
   +   n q^{\al\bt} \dl_{n+m,0}\  k,
\EE
where

\BL{}
   f_{\pm}^{0\pm}   =\pm 2\,,\ f_0^{+-}=1\,,\quad q^{00}=2\,,
   \  q^{+-}=1=q^{-+}.
\EE

The decoupling of the null vector translates via the
Ward identities to an equation
for the $n$-point functions. In the simplest case $2j'=1, j=0$,
i.e., $2J+1 =K+1\ $, and hence $m=1$; when applied to the
4-point function (for $2j'_n=1$)
the equation corresponding to (\ref{s1})
leads to a relation for the quasithermal integrals
$I_t, I_{t+1}, I_{t+2}\,$ for any $t\ge 0$.
It coincides exactly with the relation
(\ref{conv}) found above using the duality arguments
and we recall that the same linear relation was shown
to hold using the explicit expressions for the set of integrals
solving the KZ equation.
One can expect that this remains true in general, i.e., the relations
resulting from duality  are equivalent in the quasithermal
case to the algebraic equations corresponding to (\ref{s1})
when $\,2j+1=1\,$.

 The next example covered by the
formulae (\ref{VV}), (\ref{s1})
is the simplest non-thermal example $\,2j=1=2j'\, $,
 i.e., $\,m_1=2J+1=2+K\,$, $\,m=2j+1=2\,$.
In this case one expects a relation involving 5 consequent
integrals in the $(x-z)$ \? expansion, in particular a relation
expressing $I_4$ in terms of $\,I_t$, $\,t=0,1,2,3$. In other words
 we would get a system of rank four, i.e., a
$4^{th}$  order equation for the minimal DF correlation
function.

Thus in general one can use one of the algebraic equations
(that corresponding to (\ref{s1}) ) in order to select the
identities which lead to the truncation of the KZ-ZF system
from above. Yet it is rather difficult technically  to do this
explicitly, since these algebraic equations
 --  being actually differential equations
with respect to the variables $x_a$ --  are very complicated.

\sect{Relations in cohomology and differential equations}

In this section we will derive cohomological relations
(the linear and recursion relations)  and
 differential equations for the integrals under consideration.

\subsect{Twisted cohomology}

We recall our convention:
$a,b$ are indices (labelling points $z_a$, $z_b$) which run
from 1 to $\,n-1\,$; $\,i,j\,$ (or $\,i',j'\,$) run from 1
to $\,s\,$ (or $\,s'\,$) and label the insertion points
$u_i$, $u_j$ of type 1 (or type 2)
screening operators. When we derive simultaneously the thermal
and quasithermal relations we will assume $\ 1\le i,j\le m\ $,
thus $\,m=s\,$ in the thermal and $\,m=s'\,$ in the
quasithermal case. Also $\sum_i$ will mean summation
from 1 to $m$ while $\sum_a$ will mean summation from 1 to
$(n-1)$. The same will apply to products. Also $\di$ and $\da$
will be partial derivatives with respect to $u_i$ and $z_a$,
respectively. Denote by $\ve_i$ the $m$ vector with
$(\ve_i)_j=\delta_{ij}$ while $\ve_a$ is an $(n-1)$ vector with
$(\ve_a)_b=\delta_{ab}$.
 Let $l=(l_i)$ and $\mu=(\mu_a)$ be $m$ and $(n-1)$ \?
vectors, respectively.

It is convenient to introduce some notation. Define
\BE
   \ppm = \prod_a \left(\sum_i\fia\right)^{\ma}
\EE
which can be written also as
\BL{psi}
   \ppm=\sum_{A\in {\cal A}_{\mu} } \frac{\mu !}{A!(u-z)^A}
\EE
where the sum is over rectangular matrices $A=(\iaa)$
from the set
$ {\cal A}_{\mu} =\{A: \sum_i\iaa=\ma, \quad a=1,\dots,n-1\}$
and we have adopted for short a multi-index notation
$\mu != \prod_a \ma!$,
$A!(u-z)^A = \prod_a\prod_i \iaa! \ia^{\iaa}$.
For the monomials in (\ref{psi}) define
$ \deg_i \prod_{i,a}\ia^{-\iaa} = \sum_a\iaa =  l_i $
and let $P_i^{l_i}$ project on functions having degree $l_i$
in $u_i$. For an $m$ vector $l=(l_i)$ define
$ P^l=\prod_i P_i^{l_i} $.
Then we have
\BE
   \flm\equiv P^l\ppm
   = \sum_{A\in {\cal A}^l_{\mu} } \frac{\mu !}{A!(u-z)^A}
\EE
where
${\cal A}^l_{\mu}=\{A\in{\cal A}_{\mu}:
 \sum_a\iaa=l_i, \qquad i=1,\dots,m\}$.
Notice that $\flm\equiv P^l\ppm$ is different from zero only if
$$
  |l|=\sum_i l_i = \sum_a \ma =|\mu| \, .
$$
Obviously these projectors satisfy
\BE
   \di\ppi=P^l\di, \qquad \fia\ppi =P^l\fia .
\EE

In (\ref{phiL}) we have introduced $\,\fLm\,$. We will also need
\BL{phiLi}
   \fLim = \frac{(L!)^m}{(mL)!}\sum_{\z\in I(r-1,m)}\fIim,
   \qquad r=Lm-|\mu| .
\EE
We will use $\fLm$ to denote both the function and the $m$-form
$\fLm\, du_1\wedge\dots\wedge du_m$ because either our
statements will apply for functions and forms or it will be clear
from the context what is assumed. The same applies for $\fLim$.
All these forms
are considered as representing classes in the twisted cohomology
with coefficients in the local system
\BE
   \FL = \prod_{i<j}\ij^{2\nu}\prod_{i,a} \ia^{2\la}.
\EE
That is to say, in place of the ``untwisted'' exterior derivative
$d=\sum_i \di du_i$ we have the covariant derivative
\BE
   \DL=\sum_i \DLi du_i, \qquad
   \DLi=\frac 1L \left(\di + \di\log\FL\right).
\EE
Let us emphasize that this local system is invariant
under permutation of the $u$'s.\footnote {The
permutation symmetry is made explicit in the forms
(\ref{phiL}) by the averaging over $\z$.}
Our notation is designed to treat simultaneously the thermal and
quasithermal cases.
Integrating the (quasi)thermal forms $\,\ff^{[L]}_{\mu}\,$
one obtains the (quasi)thermal integrals
\BL{Int}
   I^{[L]}_{\mu}(z)=\int\Phi^{[L]}(u,z)\,\ff^{[L]}_{\mu}(u,z).
\EE
In the thermal case we have
\BE
   L=1,\quad \nu={1\over k+2},\quad \la={-J_a\over k+2},\quad m=s
\EE
and obviously
$\Phi^{[1]}\equiv \Phi^{(\nu)}_{J}(u,z)\equiv
\Phi^{(s,0;\nu)}_{J}$.
In the quasithermal case
\BE
   L=K\equiv -(k+2),\quad \nu=k+2,\quad\la=J_a,\quad m=s'
\EE
and the screening variables in this case are $v_{i'}$ instead of
$u_i$ with $i'=1,\dots,s'$. The identification
$\Phi^{[K]}\equiv \Phi^{(\nu)}_{-J /\nu}(v,z)\equiv
\Phi^{(0,s';\nu^{-1})}_{J}$ is again obvious.
The flat connection coming from the local system can be written in
a unified way for the thermal and quasithermal cases
\BL{dlogPhi}
   \frac 1L \di \log\FL = \frac 1K
   \left(\sum_a\frac{2J_a}{\ia}-2L\sum_{j(\ne i)}\fij\right)
\EE

The non-thermal local system was also introduced:
$\Phi^{(s,s';\nu)}_{J}=\Phi^{[1]}\Phi^{[K]}
\prod_{i,i'} (u_i-v_{i'})^{-2}$. The coboundary operator
is
\BE
   D = D^{[1]} + D^{[K]}
\EE
with
$D^{[1]}=\sum_i D^{[1]}_i\ du_i$ where
\BE
   D^{[1]}_i=(\di+\di\log\Phi^{(s,s';\nu)}_{J})=
   \di+\di\log\Phi^{[1]}-2\sum_{i'}(u_i-v_{i'})^{-1}
\EE
and
$D^{[K]}=\sum_{i'} D^{[K]}_{i'}\ dv_{i'}$ where
\BE
   D^{[K]}_{i'}=\frac 1K
   \left(\dz_{i'}+\dz_{i'}\log\Phi^{(s,s';\nu)}_{J}\right)
   =\frac 1K \left(\dz_{i'}+\dz_{i'}\log\Phi^{[K]}
   +2\sum_i (u_i-v_{i'})^{-1}\right)
\EE

In the next subsection we will derive relations
which are equivalent to the following cohomological relation
($\prod_{j(\neq i)}du_j$ is shorthand for
$(-1)^{i+1} du_1\wedge\dots du_{i-1}\wedge du_{i+1}\dots du_m$)
\BL{A1}
  \DL \sum_i \fLim\ \prod_{j(\neq i)}du_j\,\,  \simeq 0
\EE
in the (quasi)thermal case and to
\BL{A1.5}
   D \left(\sum_i\fsst\,\prod_{j(\neq i)}du_j\prod_{j'}dv_{j'}+
   \sum_{i'}\fssq\,\prod_j du_j\prod_{j'(\neq i')}dv_{j'}
   \right) \simeq 0
\EE
in the general case.
The form $\,\fss\,$ was introduced in (\ref{nphi}) while
$\,\fsst\,$ (or $\,\fssq\,$) are obtained if in (\ref{nphi})
one replaces $\,\ff^{[1]}_{\tau}\,$ by  $\,\ff^{[1],i}_{\tau}\,$
(or  $\,\ff^{[K]}_{\sg -\tau}\,$ by $\,\ff^{[K],i'}_{\sg -\tau}\,$).
Here with $\simeq$ we denote
equivalence in cohomology.\footnote{
All our cohomological relations are for the top forms,
i.e., $m$-forms in the (quasi)thermal case and $(s+s')$-forms
in the non-thermal case.}
 The recursion relations
which we will derive
in subsection 3 are equivalent to
\BL{R1}
  \DL \sum_i u_i \fLim\,\,\prod_{j(\neq i)}du_j\,\,\simeq 0
\EE
for the (quasi)thermal case and
\BL{R1.5}
   D \left(\sum_i u_i\ \fsst\,
   \prod_{j(\neq i)}du_j\prod_{j'}dv_{j'} +
   \sum_{i'}v_{i'}\,\fssq\,\prod_j du_j\prod_{j'(\neq i')}dv_{j'}
   \right)\simeq 0
\EE
in the general case.
Since the hypergeometric integrals are assumed to be over closed (in the
twisted homology) contours (cycles), the cohomological
identities for the forms turn into ordinary identities for the integrals.

The derivations in the next subsections are very technical
but basically they will involve the above cohomological
identities, partial fraction expansions and the following formulae:
\BL{B}
   \sum_a\ma P^l\fia\pma=l_iP^l\ppm
\EE
or
\BL{B'}
   \sum_a \frac{\ma}{\ia}\ \varphi^{[L],i}_{\mu-\ve_a}
   = L\ \fLm - (Lm-|\mu|)\ \fLim\ .
\EE
To prove the first one, notice that the l.h.s. can be written as
$$
  \sum_a\sum_{A\in{\cal A}^l_{\mu}}\iaa\frac{\mu!}{A! (u-z)^A}.
$$
Taking first the sum over $a$ we obtain the r.h.s. of (\ref{B}).
To prove the second formula we should average over $l$ in
(\ref{B}) and notice that
\BL{b}
  \sum_{\z\in\z (r,m)} (\zz)_i\ \fIm = r\ \fLim.
\EE

\subsect{Linear relations}

The (quasi)thermal linear relation is
\BL{A2}
   \sum_a (2J_a-\ma) \fLma \simeq 0
\EE
or, after integrating,
\BL{A3}
   \sum_a (2J_a-\ma) I^{[L]}_{\mu+\ve_a} = 0.
\EE
As we pointed out, these relations follow from the cohomological
identity (\ref{A1}). In order to prove (\ref{A2}) we will
establish the following identity:
\BL{aa1}
   \sum_i \di \fLim = -\sum_a \ma\,\fLma + 2L\, \sij \fij\fLim\ .
\EE

In the quasithermal case the relation (\ref{A2}) follows
immediately from (\ref{A1}) and (\ref{aa1}) (recall the flat
connection given in (\ref{dlogPhi})). In the thermal case
notice that the l.h.s. of (\ref{aa1}) is zero (becase
$\ff^{[1],i}_{\mu}$ does not depend on $u_i$). Eliminating the
term with $\sij \ij^{\,\,\,-1}...$ from (\ref{aa1}) and
(\ref{dlogPhi}) we obtain (\ref{A2}).

Now we will derive (\ref{aa1}).
First let us show
\BL{aa2}
  \sum_i\di\ppi\ppm =-\sum_a \ma P^l\ppa +2\sij\frac{l_j}{\ij}\ppi\ppm .
\EE
Indeed, differentiating we have
$$
  \sum_i P^l\di\ppm=\sum_a \ma P^l \left(-\ppa+\pma\sij\frac
  1{\ia\ja}\right).
$$
The second term on the r.h.s. can be written as
$$
  -\sij\fij\sum_a\ma\left(\fia-\fja\right)P^{l-\ve_i-\ve_j}\pma.
$$
Applying formula (\ref{B}) for the sum over $a$ we arrive
at (\ref{aa2}). Next we take the average over $l$ in (\ref{aa2}),
i.e.,  write $l=L\rho -\zz$ and sum over $\z \in \z (r-1,m)$,
$r=Lm-|\mu|$. The second term on the r.h.s. of (\ref{aa2})
gives
$$
  2L\sij \frac 1{\ij} \fLim - 2\sij \sum_{\z \in \z (r-1,m)}
  \frac {(\zz)_i} {\ij} \fIim   .
$$
The second term in the above is zero because
taking the sum over $\z$ , in a way similar to (\ref{b}),
 we end up with a
symmetric sum over an antisymmetric quantity
$$
  \sij \fij \ff^{\dots -\ve_i -\ve_j}_{\mu} \dots = 0.
$$
Thus formula (\ref{aa1}) is proved.

Now we proceed to prove the general non-thermal relations
\BL{A4}
\sum_a\ (2J_a-\sg_a)\ \fssa \simeq 0
\EE
or
\BL{A44}
\sum_a\ (2J_a-\sg_a)\ I^{(s,s')}_{\sg+\ve_a} = 0
\EE
The derivation is not difficult but long, and we will
shorten it at the cost of even greater proliferation
of notation; thus let us set temporarily:
\begin{eqnarray*}
  {\cal T}&\equiv&\sum_r {R-1\choose r}\sumam\chosa
  \sumi\fiiq\ff^{[1],i}_{\al}\ff^{[K]}_{\sg-\al}
\\ \\
  {\cal T'}&\equiv&\sum_r {R-1\choose r}\suma\chosa
  \sumi\fiiq\ff^{[1]}_{\al}\ff^{[K],i'}_{\sg-\al}
\\ \\
  {\cal X}&\equiv&\sum_r {R-1\choose r}\sumam\chosa\sum_a
  (2J_a-\al_a)\
  \ff^{[1]}_{\al+\ve_a}\ff^{[K]}_{\sg-\al}\ ,
\\ \\
  {\cal X'}&\equiv&\sum_r {R-1\choose r}\suma\chosa\sum_a
  (2J_a-\sg_a+\al_a)\
  \ff^{[1]}_{\al}\ff^{[K]}_{\sg-\al+\ve_a}\ ,
\end{eqnarray*}
$$
  {\cal Z}\equiv\sum_r {R-1\choose r}\sumam\chosa\sum_a
  (\sg_a-\al_a)\ \ff^{[1]}_{\al+\ve_a}\ff^{[K]}_{\sg-\al}\ =
  \sum_r {R-1\choose r} \suma\chosa\sum_a \al_a \
  \ff^{[1]}_{\al}\ff^{[K]}_{\sg-\al+\ve_a}\ .
$$
{}From (\ref{A1.5}) we have the cohomological identity
(derived in the same way as the (quasi)thermal relations):
\BL{aa6}
  {\cal X}+{\cal X}' \simeq 2\ (K\ {\cal T}-{\cal T}')\,.
\EE
We obtain $(K\ {\cal T}-{\cal T}')$ if we apply the basic
identity (\ref{B'}) to
\BL{aa7}
  \sum_{r;i,i';a} {R-1\choose r} \fiiq \bigr(
  \sumam\chosa \frac{(\sg_a-\al_a)}{\iqa}
  \ff^{[1],i}_{\al}\ff^{[K],i'}_{\sg-\al-\ve_a}
  - \suma\chosa\frac{\al_a}{\ia}
  \ff^{[1],i}_{\al-\ve_a}\ff^{[K],i'}_{\sg-\al}\bigl)\ .
\EE
On the other hand, using
$$
  \frac 1{\iiq\iqa}-\frac 1{\iiq\ia}=\frac 1{\ia\iqa}
$$
we obtain that (\ref{aa7}) is equal to ${\cal Z}$.

Using a standard binomial coefficient formula
$\chosa+{\sg\choose\al-\ve_a}={\sg+\ve_a\choose\al}$
in a multi-index guise we get from (\ref{aa6}):
$$
  ({\cal X}-{\cal Z})+({\cal X}'-{\cal Z}) =
  \sum_a(2J_a-\sg_a)\fssa \simeq 0\ .
$$
i.e., the general non-thermal Aomoto-type relation.

\subsect{Recursion relations}

The (quasi)thermal recursion relation is
\BL{R2}
   \sum_a (2J_a -\ma)\ z_a\ \fLma
   + \, c^{[L]}_{|\mu|}\ \fLm  \simeq 0
\EE
where
\BL{c}
   c^{[L]}_r = 2\ \sum_{a=1}^{n-1} J_a - r - L\,m+1+K
\EE
Of course, integrating (\ref{R2}) we obtain the recursion
relation for the respective integrals. In order to prove the
recursion relation we will have to establish the following identity
\BL{rr1}
   \sum_i u_i\ \di\,\fLim = -\sum_a \ma\, z_a\, \fLma -
   (|\mu|+(m-1)L)\ \fLm
   +2\,L\ \sij\frac{u_i}{\ij}\ \fLim\,.
\EE
Note that
\BL{rr1.5}
   \frac KL \sum_i u_i\, \fLim\, \di\log\FL =
   \sum_a 2J_a\, \fLm +\sum_a 2J_a\,z_a\,\fLma -
   2\,L\ \sij\frac{u_i}{\ij}\ \fLim\,.
\EE
Therefore in the quasithermal case the cohomological
identity (\ref{R1}) immediately translates into the
recursion relation (\ref{R2}) when we combine (\ref{rr1})
and (\ref{rr1.5}). In the thermal case notice that the
l.h.s. of (\ref{rr1}) vanishes. Thus we can eliminate
the last terms of (\ref{rr1}) and (\ref{rr1.5}) and
again (\ref{R1}) gives us (\ref{R2}).

Now we proceed to prove (\ref{rr1}). From
the direct differentiation we have
\BAN
   \lefteqn{ \sum_iu_i\di \ppi\ppm \,\,=\,\,
   \sum_i\sum_a(z_a+\ia)\ma P^l(\pma\di \psi_a) = }
\\ \nonumber \\
   &-&\sum_a\ma z_a P^l \pma\left(\psi_a^2 -\sij\frac
  1{\ia\ja}\right) - \sum_i\sum_a\ma\ppi\frac{\pma}{\ia} =
\nonumber  \\ \nonumber   \\
   &-& \sum_a\ma z_a P^l\ppa + \sum_a \ma z_a P^l \ppa
   \sij\frac 1{\ia\ja}  - \sum_i (l_i-1)\ppi \ppm
\label{rr2}
\EEA
where to obtain the third term of (\ref{rr2}) we have used
(\ref{B}). Substitute $z_a=-(u_i-z_a)+u_i$, in  the
second term of (\ref{rr2}) and rewrite this term
as follows
$$
  -\sij\sum_a\ma\ppi\frac{\pma}{\ja} + \sij\frac{u_i}{\ij}
  \sum_a\ma\left(\fja-\fia\right)P^{l-\ve_i-\ve_j}\pma.
$$
Applying (\ref{B}) to the above we get
\BL{rr3}
   -\sij l_j \ppi \ppm + 2\sij\frac{u_i}{\ij}
   \left(l_j\ppi-l_iP^{l-\ve_j}\right)\ppm =
   -2\sij l_j\ppi\ppm +2\sij\frac{u_i}{\ij}l_j\ppi\ppm.
\EE
Now substitute back (\ref{rr3}) for the second term of (\ref{rr2}).
and rewrite $-\!\sum_i(l_i-1)\ppi -2\sij l_j \ppi $ as
$-\!|\mu|\sum_i\ppi-\sij l_j \ppi $,
taking into account that  $|\mu|=|l|-1$ in this case.
Thus we obtain the following identity:
\BL{rr4}
   \sum_iu_i\di \ppi\ppm = -\sum_a\ma z_a P^l \ppa
   -|\mu|\ \sum_i\ppi\ppm   +
   \sij\left(\frac{2\ u_i}{\ij}-1\right) l_j\ \ppi\ppm\  .
\EE
If we perform the averaging over $l$, i.e., set
$l=L\rho -\zz$ and sum over $\z$ we arrive at
 (\ref{rr1}) and thus we have derived
the (quasi)thermal recursion relations.

The proof of the general recursion relation
\BL{R3}
   \sum_a(2J_a-\sg_a)\ z_a\ \fssa
    + c_{|\sg|}\ \fss \simeq 0 ,
\EE
where
\BL{c1}
   c_{|\sg|} \equiv 2J_n+1+K+S-|\sg|
   = \sum_{a=1}^n J_a +1+K-|\sg|
   = 2\sum_{a=1}^{n-1} J_a +1+K-S-|\sg|\, ,
\EE
though somewhat more involved, is nevertheless analogous
to the derivation of the general non-thermal linear relation
in the previous subsection and we will skip it.

\subsect{Differential equations}

In this section we will show that the integrals (\ref{Int}) satisfy
the following system of
differential equations
\BA{GM}
   \lefteqn{ -K\ \da\ I^{[L]}_{\mu} =
   (Lm-|\mu|)\, (2J_a-\ma)\, I^{[L]}_{\mu+\ve_a} }
\\
   && -\sba \frac 1{(z_a-z_b)} \left[
   \ma(2J_b-\mb)(I^{[L]}_{\mu}-
   I^{[L]}_{\mu-\ve_a+\ve_b})+\mb(2J_a-\ma)(I^{[L]}_{\mu}
   - I^{[L]}_{\mu+\ve_a-\ve_b})\right]  \nonumber
\EEA
or, written in terms of a cohomological identity for
$m$-forms:
\BA{GM2}
   \lefteqn{ -K\ D_a\ \ff^{[L]}_{\mu} \simeq
   (Lm-|\mu|)\, (2J_a-\ma)\, \ff^{[L]}_{\mu+\ve_a} }
\\
   && -\sba \sum_i \frac 1{\ia\ib} \left[
   \ma(2J_b-\mb) \ff^{[L],i}_{\mu-\ve_a}
   -\mb(2J_a-\ma)
    \ff^{[L],i}_{\mu-\ve_b})\right]  \nonumber
\EEA
where $D_a=\da + \da\log\FL$.

Before we proceed let us introduce the notation
$$
  {\cal M}_{ab} = \sum_i \frac 1{\ia\ib}
  \ff^{[L],i}_{\mu-\ve_a} .
$$
If $a\neq b$ we can rewrite it as follows
$$
  {\cal M}_{ab} =  \frac 1{(z_a-z_b)}
  (\ff^{[L]}_{\mu}- \ff^{[L]}_{\mu-\ve_a+\ve_b})\,,
$$
and in particular we see that integrating (\ref{GM2})
we obtain (\ref{GM}). One immediately obtains
(\ref{GM2}) if the following identities are established
(recall that we denote $\psi_a=\sum_i (\ia)^{-1}$):
\BA{gm1}
   && -K\, D_a\, \fLm = 2J_a\, L\, \psi_a\fLm
     - \ma\, K\, {\cal M}_{aa}
\\ \nonumber\\ \label{gm2}
   && \sum_b \mb\, {\cal M}_{ba} - L\, \psi_a\fLm
      + (Lm-|\mu|)\, \fLma = 0
\\ \label{gm3}
   && \sum_b(2J_b-\mb)\, {\cal M}_{ab} +
       L\, \psi_a\fLm - K\, {\cal M}_{aa} \simeq 0 .
\EEA
The first equation follows directly by differentiating,
the second follows immediately from our basic relation
(\ref{B'}) while the cohomological identity (\ref{gm3})
we will derive from the cohomological identity
\BL{GM1}
   D \sum_i \fia \ff^{[L],i}_{\mu-\ve_a}
   \prod_{j(\neq i)} du_j\simeq 0 \, .
\EE
To prove (\ref{gm3}) we will establish
\BL{gm4}
   \sum_i \fia \di \ff^{[L],i}_{\mu-\ve_a} =
   -\sum_b (\mu-\ve_a)_b\ {\cal M}_{ab} +
   L \sij \fij \left(\fia+\fja\right)
   \ff^{[L],i}_{\mu-\ve_a}.
\EE
In the thermal case the l.h.s. of (\ref{gm4}) is zero, and
adding (\ref{gm4}) multiplied by $K^{-1}$ to (\ref{GM1})
we obtain (\ref{gm3}). In the quasithermal case,
substituting (\ref{gm4}) into (\ref{GM1}) we obtain (\ref{gm3}).
So let us now derive (\ref{gm4}). Differentiating we
obtain
\begin{eqnarray*}
  \lefteqn{ \sum_i \fia P^l \di \pma =
  - \sum_i\sum_b \frac{(\mu-\ve_a)_b}{\ia\ib}\ppi\pma }
\\
  && +\sij\sum_b \frac{(\mu-\ve_a)_b}{\ij}
  \left(\frac 1{\jb} - \fib\right)
  \ppi \frac{\psi^{\mu-\ve_a-\ve_b}}{\ia}\,.
\end{eqnarray*}
Apply to the second term of the r.h.s. our identity
(\ref{B}) to obtain
$$
  \sum_i \fia \di \ff^{l-\ve_i}_{\mu-\ve_a} =
  - \sum_b \sum_i \frac{(\mu-\ve_a)_b}{\ij}
  \ff^{l-\ve_i}_{\mu-\ve_a}
  +\sij \frac{l_j}{\ij}
  \left(\frac 1{\jb} + \fib\right) \ff^{l-\ve_i}_{\mu-\ve_a}\ .
$$
Now average over $l$ in the above. As in the previous
cases, for part of the second term we get a zero
because of a symmetric sum over an antisymmetric
quantity and this completes the derivation of (\ref{gm4})
and, as already discussed,  also of (\ref{GM2}).

The general non-thermal differential equation is
\BAN
   \lefteqn{ -K\,D_a\,\iiss_{\sg}
   =(S-|\sg|)\,(2J_a-\sg_a)\, \iiss_{\sg+\ea} }
\\ \label{GM3}  \\ \nonumber
   &-&\sba \, {1 \over \zab} \left(
   \sg_a\,(2J_b-\sg_b)\,(\iiss_{\sg}-\iiss_{\sg-\ea+\eb}) +
   \sg_b\,(2J_a-\sg_a)\,(\iiss_{\sg}-\iiss_{\sg+\ea-\eb})\right).
\EEA

Its derivation is again similar to the derivation of the general
non-thermal linear and recursion relations.
Using the complete non-thermal local system we obtain, in analogy with
the (quasi)thermal case (\ref{GM2}), the following cohomological
identity
\BAN
   && -K\,D_a\, \phi_{\al,\bt} \simeq
    2\,\left( K\,\ala\, {\cal T}_{\al-\ea,\bt}
           - \bta\, {\cal T}'_{\al,\bt-\ea} \right)
    + r\,(2J_a-\ala)\, \phi_{\al+\ea,\bt}+
    (R-r)\,(2J_a-\bta)\,\phi_{\al,\bt+\ea}
\\ \label{GM4} \\ \nonumber
    && - \sba \left( \ala\,(2J_b-\alb)\,{\cal M}_{ab} -
       \alb\,(2J_a-\ala)\,{\cal M}_{ba} \right)
   - \sba \left( \bta\,(2J_b-\btb)\,{\cal M}'_{ab} -
       \btb\,(2J_a-\bta)\,{\cal M}'_{ba} \right)
\EEA
where $\al+\bt=\sg$ and this time we have denoted
\BAN
   {\cal M}_{ab}={1\over\zab}\,(\ff^{[1]}_{\al}\ff^{[K]}_{\bt}
      -\ff^{[1]}_{\al-\ea+\eb}\ff^{[K]}_{\bt}),
   &\quad&
   {\cal M}'_{ab}={1\over\zab}(\ff^{[1]}_{\al}\ff^{[K]}_{\bt}
      -\ff^{[1]}_{\al}\ff^{[K]}_{\bt-\ea+\eb}),
\\ \\ \nonumber
   {\cal T}_{\al-\ea,\bt}=\sumi\,{1\over\iiq\ia}\,
    \ff^{[1],i}_{\al-\ea}\ff^{[K]}_{\bt},
   &\quad&
   {\cal T}_{\al,\bt-\ea}=\sumi{1\over\iiq\iqa}
    \ff^{[1]}_{\al}\ff^{[K],i'}_{\bt-\ea}
\EEA
and again $|\al|=s-r$, and $R=S-|\sg|=Ks'-|\bt|+r$.
To compare with (\ref{GM3}) we have to multiply (\ref{GM4}) by
$\chosa$ and sum over $\al$. To obtain the
term $2(K\,{\cal T}-{\cal T}')$ of (\ref{GM4}) apply the basic identity
(\ref{B'}) to
\BE
   \sum_{\al} {S-|\sg|\choose s-|\al|}\chosa \sum_{ii';b}
   {1\over\iiq}\left( {\ala\btb
   \,\ff^{[1],i}_{\al-\ea}\ff^{[K],i'}_{\bt-\eb}\over\iqb\ia}
   - {\alb\bta\,\ff^{[1],i}_{\al-\eb}\ff^{[K],i'}_{\bt-\ea}
    \over\ib\iqa}\right) \, .
\EE
Substituting in the above the following identity,
$$
   {1\over\iiq\ia\iqb} - {1\over\iiq\iqa\ib}
$$

$$
   ={1\over\zab}\left({-1\over\ia\iqb}+{-1\over\ib\iqa}+
   {1\over\ia\iqa}+{1\over\ib\iqb}\right)
$$
and after some algebra we obtain (\ref{GM3}).

\sect{Discussion}

We have found solutions of the KZ equations of the $A_1^{(1)}$
WZNW conformal theory for arbitrary level $k\neq -2$ and some isospin
values corresponding to infinite dimensional representations
of $A_1$ --  namely the isospins of the type in (\ref{kJ}), $J_{j,j'}, \
2j,2j' \in Z\!\!\!Z_+\ $ --
although we
were mainly interested
in the  ``minimal subset'' described by rational $k$ and
by the domain in (\ref{rat}), (\ref{jj'}).

These are precisely the values which parametrize  according to
(\ref{c_k}), (\ref{h(J)-J}), all reducible Virasoro Verma
modules. Reducibility implies differential (BPZ) equations
for the conformal blocks of primary fields. Our results show
that these blocks can be ``imbedded'' in a set of generalized
hypergeometric integrals which satisfy a first-order linear
differential system of equations and serve as coefficients
of the infinite $\ (x-z)$ \? expansion representing the
correlators of the $A^{(1)}_1$ WZNW model. We have given
strong arguments that the KZ equation, written as an infinite --
in general -- system for the coefficients in the
$(x-z)$ \? expansion, reduces
to a finite system from which one can recover the BPZ equation for
the Virasoro correlator. Yet it
remains a rather difficult technical problem to do this explicitly in
general -- either by exploiting the existing finite system in \cite{A}, or
by taking into account that the solutions of the KZ equation satisfy
 also $A_1^{(1)}$ null vector decoupling  equations.

This is related to the problem of constructing the
expressions for the singular vectors
of the Virasoro  Verma modules ``via quantum Hamiltonian reduction''.
The simpler thermal (and quasithermal) cases discussed above shed some
light on the relation of our construction and the approach in \cite{BFIZ}
where an algorithm for obtaining all Virasoro singular vectors is
proposed (see also the recent paper \cite{BS}, which we received while
this work was in process.)
  In particular in  \cite{BFIZ} the analogue of our infinite
system in the general non-thermal case is avoided, using the fusion
procedure.

The peculiarities of the rational level cases were only
briefly touched in this paper. The rational level theories require a further
and more thorough study -- especially in connection with the description of
the relevant twisted homology and cohomology groups.

In our ``$(x-z)$'' \? expansion of the correlators and the proposed
solution  in terms of ``meromorphic'' forms we have departed
from bosonization. This raises the problem
of finding a proper operator language behind the construction.

There is an important open problem which we have not touched
so far, namely the understanding of the isospin
$sl_I(2,C\!\!\!\!I)$
tensor product decomposition of the  correlation functions.
This requires yet another expansion,  extending  the representation
 used in
\cite{ChF} to the case of non-integer isospins $2J_a$.

Finally let us point out that the generalization of our
results to the case of higher rank algebras is also
possible.

\vskip 1cm
\noindent{\bf Acknowledgements}
\vskip .5cm

The authors would like to thank
L. Alvarez-Gaum\'{e},  P. Bouwknegt, L. D\c{a}browski, V.K. Dobrev,
Vl.S.  Dotsenko, V.G. Kac,
and C. Reina  for useful discussions.
V.B.P. acknowledges the hospitality and the financial
support of INFN and SISSA.

\vskip 1cm
\renewcommand{\theequation} {A.\arabic{equation}}
\appendix
\sect{Appendix}

Here we will establish an algorithm determining $\BBB$.
The notation is the same as in Section 4.2.
Consider  the (+1), (0), ($-1$), $\dots$ diagonals of (\ref{BF=FB}).
{}From this equation we can determine  successively
 the matrix elements of
$\BBB^{(0)},\BBB^{(-1)},\BBB^{(-2)},\dots\ $.

First consider the (+1) diagonal of (\ref{BF=FB}).
We have $ (\BBB^{(0)}\FFF^{(1)})_{ii+1}
=(\hat \FFF^{(1)}\BBB^{(0)})_{ii+1} $ or
$$
  \BBB^{(0)}_{i+1}
  =\BBB^{(0)}_i\ \FFF^{(1)}_i/\hat \FFF^{(1)}_i
$$
while from (\ref{I0=I0}) we have
$$
  \BBB^{(0)}_0=1
$$
thus
\BE
  \BBB^{(0)}_{i+1}=\prod _{k=0}^i
  \left( \FFF^{(1)}_k\over\hat \FFF^{(1)}_k\right)
  \qquad i=0,1,2\dots
\EE
and all $\BBB^{(0)}_i$ are constants
(i.e., no $z$ dependence).

{}From the $(-n+1)$ diagonal of (\ref{BF=FB})
we get a simple recursion for the elements of $\BBB^{(-n)}$
of the form:
\BL{BCD}
   \BBB^{(-n)}_{j+1}=\DD^{(n)}_j\BBB^{(-n)}_j+\CC^{(n)}_j .
\EE
The initial condition
$$\BBB^{(-n)}_{n-1}=0$$
is obvious if we note that the indices of our
matrices $\BBB$ and $\FFF$ run over the non-negative
integers. The solution of this recursion  is
$$
  \BBB^{(-n)}_{j+1}=\CC^{(n)}_j+ \sum _{i=0}^{j-n}
  \DD^{(n)}_j\DD^{(n)}_{j-1}\dots \DD^{(n)}_{j-i}
  \CC^{(n)}_{j-i-1}, \quad j=n,n+1,n+2,\dots .
$$

Let us consider the first few values of $n$.
The (0) diagonal of (\ref{BF=FB}) gives:
$$
  \BBB^{(0)}(\FFF^{(0)}+\partial)+\BBB^{(-1)}\FFF^{(1)}=
  (\hat \FFF^{(0)}+\partial)\BBB^{(0)}
  +\hat \FFF^{(1)}\BBB^{(-1)}\,.
$$
Because $\BBB^{(0)}$ is a constant matrix, the terms with $\partial$
cancel. We get a recursion of the form (\ref{BCD}) with
$\DD^{(1)}_i=\FFF^{(1)}_{i-1} /\hat \FFF^{(1)}_i$ and
$\CC^{(1)}_i= \BBB^{(0)}_i
(\FFF^{(0)}_i-\hat \FFF^{(0)}_i)/\hat \FFF^{(1)}_i$
and initial condition
$\BBB^{(-1)}_0=0 $
(it is easy to note that in fact we have $\BBB^{(-1)}_1=0$).

{}From the $(-1)$ diagonal  of (\ref{BF=FB}) we obtain:
$$
 \BBB^{(0)}\FFF^{(-1)}+\BBB^{(-1)}\FFF^{(0)}+\BBB^{(-2)}\FFF^{(1)}
 =\hat \FFF^{(-1)}\BBB^{(0)}+\hat \FFF^{(0)}\BBB^{(-1)}+
 (\partial \BBB^{(-1)})+\hat \FFF^{(1)}\BBB^{(-2)}
$$
which gives a recursion for $\BBB^{(-2)}$ with
$$
  \DD^{(2)}_i={\FFF^{(1)}_{i-2}\over\hat\FFF^{(1)}_i} ,
  \qquad
  \CC^{(2)}_i={1\over \hat \FFF^{(1)}_i}\left[
  \BBB^{(0)}_i\FFF^{(-1)}_i-\BBB^{(0)}_{i-1}\hat \FFF^{(-1)}_i
  +\BBB^{(-1)}_i(\FFF^{(0)}_{i-1}-\hat \FFF^{(0)}_i)-(\partial
  \BBB^{(-1)}_i)\right] .
$$
Obviously this is the most general case, i.e. for
$\BBB^{(-k)}$ with $k=2,3,\dots$ we have
$$
  \DD^{(k)}_i={\FFF^{(1)}_{i-k}\over\hat\FFF^{(1)}_i} ,
  \quad
  \CC^{(k)}_i={1\over \hat \FFF^{(1)}_i}\left[
  \BBB^{(-k+2)}_i\FFF^{(-1)}_{i-k+2}
  -\BBB^{(-k+2)}_{i-1}\hat \FFF^{(-1)}_i
  +\BBB^{(-k+1)}_i(\FFF^{(0)}_{i-k+1}-\hat \FFF^{(0)}_i)-
  (\partial \BBB^{(-k+1)}_i)\right] .
$$
Thus we have an algorithm to determine the matrix $\BBB$.

Let us give the answer for the first few entries:
\begin{eqnarray*}
 \BBB^{(0)}_1&=&0\,, \qquad \BBB^{(0)}_i=
 {\FFF^{(1)}_{i-1}\dots \FFF^{(1)}_{0}\over
 \hat \FFF^{(1)}_{i-1}\dots\hat \FFF^{(1)}_0}
 \quad i=1,2,\dots\,,
\\
 \BBB^{(-1)}_1&=&0\,, \qquad  \BBB^{(-1)}_2
 ={\FFF^{(1)}_0(\FFF^{(0)}_1-\hat \FFF^{(0)}_1)
 \over \hat \FFF^{(1)}_1 \hat \FFF^{(1)}_0}\,,
\\
 \BBB^{(-2)}_2&=&{\FFF^{(1)}_0\FFF^{(-1)}_1-
 \hat \FFF^{(-1)}_1\hat \FFF^{(1)}_0
 \over \hat \FFF^{(1)}_1\hat \FFF^{(1)}_0}\,,
\\
 \BBB^{(-2)}_3&=& {\FFF^{(1)}_0\over \hat \FFF^{(1)}_2\hat
 \FFF^{(1)}_1\hat \FFF^{(1)}_0}
  \left(\FFF^{(1)}_0\FFF^{(-1)}_1-\hat \FFF^{(1)}_0\hat
  \FFF^{(-1)}_1+\FFF^{(1)}_1\FFF^{(-1)}_2\right.
\\ \\
 && \left. -\hat \FFF^{(1)}_1\hat \FFF^{(-1)}_2 +
   (\FFF^{(0)}_1-\hat \FFF^{(0)}_1)(\FFF^{(0)}_1-
   \hat \FFF^{(0)}_2)
   -\partial(\FFF^{(0)}_1-\hat \FFF^{(0)}_1)\right)\,.
\end{eqnarray*}
And at the end we write down  some very explicit formulas
\BA{5b1}
   \BBB^{(0)}_2&=&{(2J_2-1)(s-1)\over (2J_2-K)(s-K)}\,,
\\ \label{5b2}
   \BBB^{(-1)}_2 &=& {(K-1)\over (2J_2-K)(s-K)}
  \left( {2J_1+2J_2 \over z}+{2J_2+2J_3\over z-1}\right)\,,
\\ \label{5b3}
  \BBB^{(-2)}_2&=&
  {(K-1)(J_1+J_2+J_3+J_4+1+K)\over(2J_2-K)(s-K)\,z(z-1)}\,.
\EEA

\end{document}